\documentclass{aa}  

\usepackage{graphicx}
\usepackage{txfonts}
\usepackage{longtable}
\usepackage{appendix}
\usepackage{caption} 
\usepackage{threeparttable} 
\usepackage{comment}
\usepackage{color}
\usepackage[dvipsnames,table]{xcolor}

\newcommand\newtext[1]{{#1\sl}}

\newcommand\gaia{\text{\it Gaia}\xspace}

\begin{document}

   \title{\gaia DR3 asteroid reflectance spectra: L-type families, memberships, and ages}

   \subtitle{Applications of \gaia spectra}

   \author{R. Balossi\inst{1}
          \and 
          P. Tanga\inst{1}
          \and 
          A. Sergeyev \inst{1, 2}
          \and
          A. Cellino \inst{3}
          \and
          F. Spoto \inst{5}
          }

   \institute{Université Côte d'Azur, Observatoire de la Côte d'Azur, CNRS, Laboratoire Lagrange, Bd de l'Observatoire, CS 34229, 06304 Nice Cedex 4, France\\
   \email{roberto.balossi@oca.eu}
         \and
             V. N. Karazin Kharkiv National University, 4 Svobody Sq., Kharkiv 61022, Ukraine
         \and
            INAF - Osservatorio Astrofisico di Torino, via Osservatorio 20, 10025, Pino Torinese (TO), Italy
        \and
            Harvard-Smithsonian Center for Astrophysics, 60 Garden St., MS 15, Cambridge, MA, 02138, USA\\
             }
   \date{}

 
  \abstract
   {The \gaia Data Release 3 (DR3) contains reflectance spectra at visible wavelengths for \newtext{60,518} asteroids over the range between 374-1034~nm, representing a large sample that is well suited to studies of asteroid families.} 
   {We want to assess the potential of \gaia spectra in identifying asteroid family members. Here, we focus on two L-type families, namely Tirela/Klumpkea and Watsonia. These families are known for their connection to Barbarian asteroids, which are potentially abundant in calcium-aluminum rich inclusions (CAIs).}
   {Our method is based (1) on a color taxonomy specifically built on \gaia data and (2) the similarity of spectra of candidate members with the template spectrum of a specific family. }
   {We identified objects in the halo of Tirela/Klumpkea, along with possible interlopers. We also found an independent group of eight asteroids erroneously linked to the family by the hierarchical clustering method (HCM). Consequently, the knowledge of the size distribution of the family has been significantly improved, with a more consistent shape at the larger end. The Watsonia family is a more intricate case, mainly due to its smaller size and the less marked difference between the spectral types of the background and of the family members. However, the spectral selection helps identify objects that were not seen by HCM, including a cluster separated from the family core by a resonance. }
   {For both families, the V-shape is better defined, leading to a revised age estimation based on \newtext{the memberships established mainly from spectral properties}. Our work demonstrates the advantage of combining the classical HCM approach to spectral properties obtained by \gaia for the study of asteroid families. Future data releases are expected to further expand the capabilities in this domain.}


   \maketitle
%

\section{Introduction} \label{Introduction}

The publication of asteroid reflectance spectra in \gaia DR3 \citep{Gaiacollaboration-2023} has been followed up by studies that exploit the large sample of more than 60,000 objects. The applications concern individual objects part of larger groups (\citealt{hasegawa-2022}, \citealt{carruba_2024}, \citealt{minker-2023}), primitive asteroids, and families (\citealt{ferrone-2023}, \citealt{delbo-2023}, \citealt{tatsumi-2023}), and near-Earth Asteroids \citep{sergeyev_2023}, as well as comparisons with andesitic meteorites \citep{galinier-2023}. While the exploitation of \gaia asteroid spectra at a high signal-to-noise ratio (S/N) appears to be generally rather straightforward, the DR3 sample poses some difficulties for fainter targets, especially concerning the behavior at the longest visible wavelengths. This is a known limitation already discussed in \newtext{\cite{Gaiacollaboration-2023}}, specific to this first large data release, which should be overcome in future publications. A less problematic, systematic bias in the extreme blue-violet region has been identified before \citep{tinautruano-2023}.

The main goal of this article is to assess the potential impact of the increased number of spectra available over family membership, thanks to \gaia DR3. This aspect involves classification attempts based on spectra alone and validation via comparisons to memberships derived by hierarchical clustering methods (HCM) alone, based on proper elements \newtext{(\citealt{zappala-1990}, \citealt{milani-2016})}. Also, albedo is exploited in the following when appropriate. We base our approach on quantitatively computing the similarity of spectra and, separately, by a classifier that groups asteroids into spectral classes. Our classifier, based on machine learning, has been optimized specifically on \gaia spectra and provides the two most probable taxonomies for each asteroid.

Two L-type families are selected for this study, namely Tirela/Klumpkea and Watsonia. They present spectral similarities and both harbor so-called Barbarians, asteroids characterized by a peculiar polarimetry (\citealt{CellinoBarbara_2006}, \citealt{gilhutton-2008}, \citealt{gilhutton-2014}, \newtext{\citealt{cellino-2014}}, \citealt{bendjoya-2022}). Their scattering properties and reflectance spectra have been associated with the possible large abundance of calcium-aluminum rich inclusions (CAIs) in a CV--like matrix, with a low degree of thermal or aqueous alteration \citep{sunshine-2008, devogele-2018}. Barbarians could represent an interesting ancient population preserving the trace of highly inhomogeneous accretion (with a preferential segregation of some highly refractory material) in the early protoplanetary disk. A recent finding that the variety of CV chondrites available could at least partially match the observed variety of Barbarians (and in general, L-type) asteroids  \citep{mahlke-2023} adds further uncertainty and interest to the interpretations about the nature of this category of objects.
In addition, the two selected families differ in several aspects, in particular, the position in the belt (middle belt for Watsonia, outer belt for Tirela); the albedo; the size distribution; and the number of known members. Thus, they offer a very different context to test our approach. Eventually, our spectra-based membership was used to re-compute the age of the family, based on the same approach as \citet{spoto-2015}.

We stress here that by adopting the spectral similarity as the main driver for our family member selection, we are enforcing the commonly accepted paradigm of compositional homogeneity within the family. The consequence, to some extent, is that we may suppress the possible evidence of any heterogeneity. However, some of the approaches we use here can also be adapted when looking for spectral diversity inside families, especially with future \gaia releases where the S/N of reflectance spectra is expected to increase for all the published asteroids.

\newtext{In the process, we compare our family memberships to those obtained by the HCM. Some cautions are due for this comparison, as several HCM implementations exist in practice and they are not all equivalent. In general, they differ by the approach used to define the cutoff distance in the proper element space (corresponding to a difference in velocity), used as a limit for linking family members. Beyond a certain distance, the ''chaining'' effect occurs (i.e., an overlap of neighboring families) and the family tends to merge into a background of small asteroids. Since, in the following, we mainly consider the HCM implementation by \citet{milani-2014, milani-2016}, we recall here that their procedure consists of choosing different quasi-random levels (defining velocity limits) for several intervals of heliocentric distance, thereby taking into account local differences in the asteroid population. Also, they divide the search for family membership into several steps of increasing distance, from family core to halo. 
It is thus clear that the details of the membership are dependent on parameters that are specific to each family identification effort. However, we do not expect these differences to substantially affect our approach, as the Gaia survey is limited in depth, in principle at absolute magnitude G$\sim$20.5, but in practice, also by the quality of the spectra. We show that our sample is incomplete below 5--8~km in diameter in the Main Belt. For the larger asteroids on which we base this work, the membership is not very sensitive to differences in the HCM implementations.}

The article is organized as follows. In Sect. \ref{Classyfing Gaia Spectra}, we briefly describe the color-based taxonomy used to classify \gaia spectra. In Sect. \ref{SectionTirela}, we present our analysis on the Tirela/Klumpkea family, focused on its membership compared to HCM and a determination of its age. In Sect. \ref{SectionWatsonia}, a similar analysis for the Watsonia family is illustrated. Finally, in Sect. \ref{Conclusion}, our results and future prospects are presented.

\section{Classifying \gaia spectra} \label{Classyfing Gaia Spectra}

\begin{figure}
   \centering
   \includegraphics[width=\hsize]{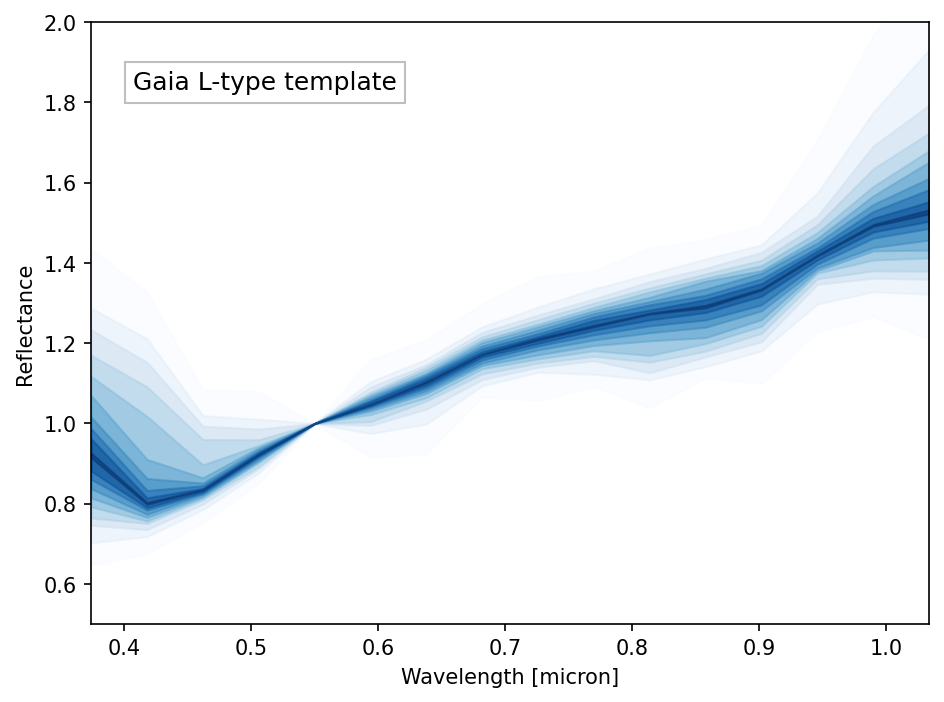}
      \caption{\small \newtext{Density plot of all the reflectance spectra of L- type asteroids observed by \gaia. Darker areas correspond to the overlap of a larger number of spectra.}}
         \label{LTypeSpectra}
\end{figure}

Widely recognized asteroid taxonomy schemes, such as those proposed by \cite{tholen-1984} and by \cite{demeo-2009}, along with the more recent framework proposed by \citet{mahlke-2022}, systematically classify asteroids based on their reflectance spectra and albedo measurements. 
The concept of asteroid taxonomy is based on analyzing the reflective properties of asteroids to determine their surface composition. This classification system leverages the spectral signatures reflected by asteroid surfaces, which indicate their compositional characteristics. 
These methods have been instrumental in linking observable traits to the surface composition of asteroids, thereby providing insights into their mineralogical makeup.

\begin{figure*}[ht]
   \centering
   \includegraphics[width=\textwidth]{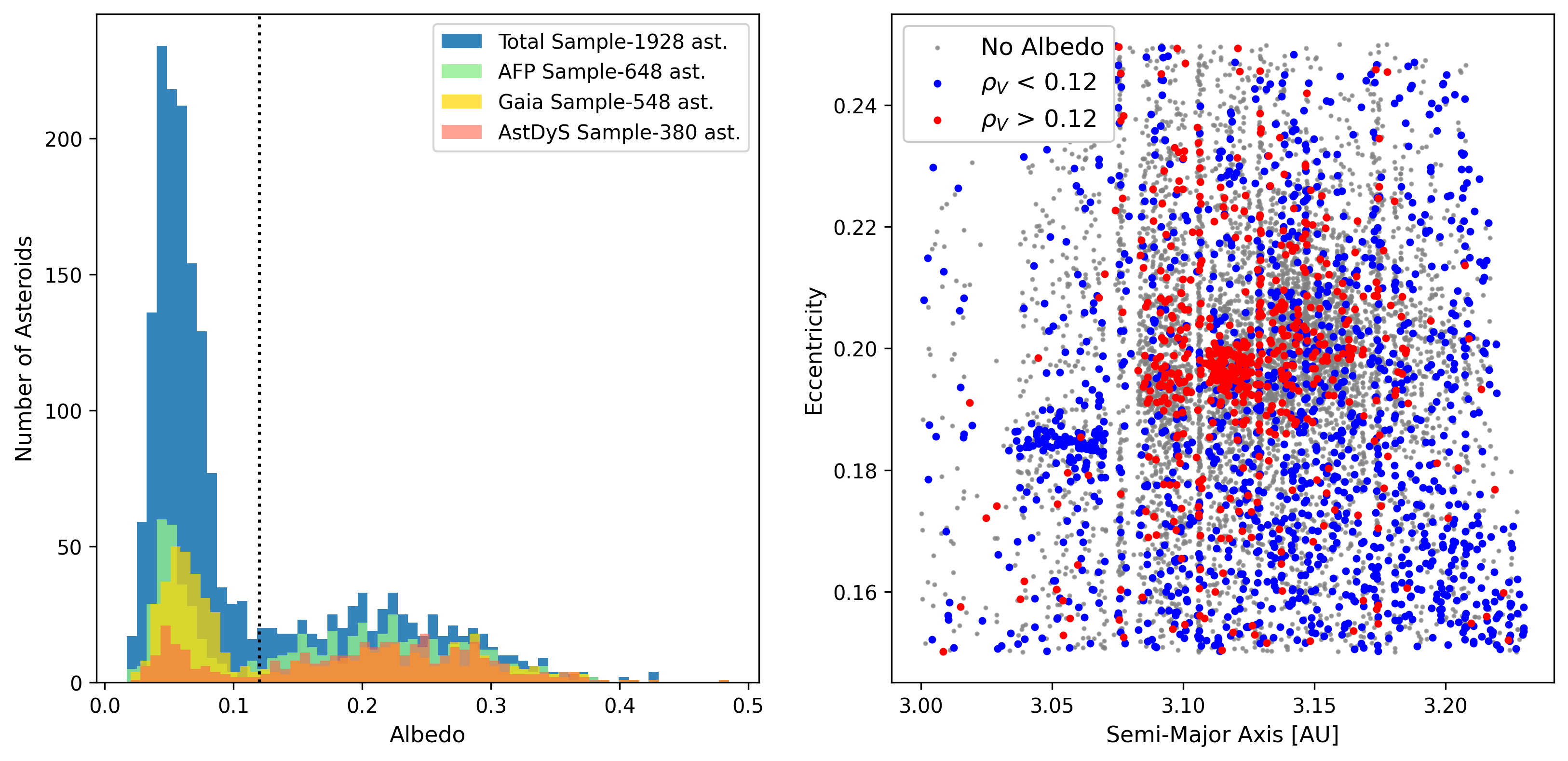}
      \caption{\small \newtext{Albedo distribution of the objects located within the Tirela/Klumpkea region.} Left panel: NEOWISE albedo distribution for the objects of the complete sample (in blue), of the \gaia sample (in yellow), of the family members listed in AstDyS (in red) and of the family members listed in AFP (in green). The vertical black dotted line at $\rho_V = 0.12$ separates low albedo from high albedo objects.
      Right panel: Distribution in the $(a, e)$ plane of the objects located within the Tirela/Klumpkea region. Blue points have $\rho_V < 0.12$, red points have $\rho_V > 0.12,$ and grey points do not have albedo measurements.}
         \label{TirelaAlbedo}
   \end{figure*}

\begin{figure}
   \centering
   \includegraphics[width=\hsize]{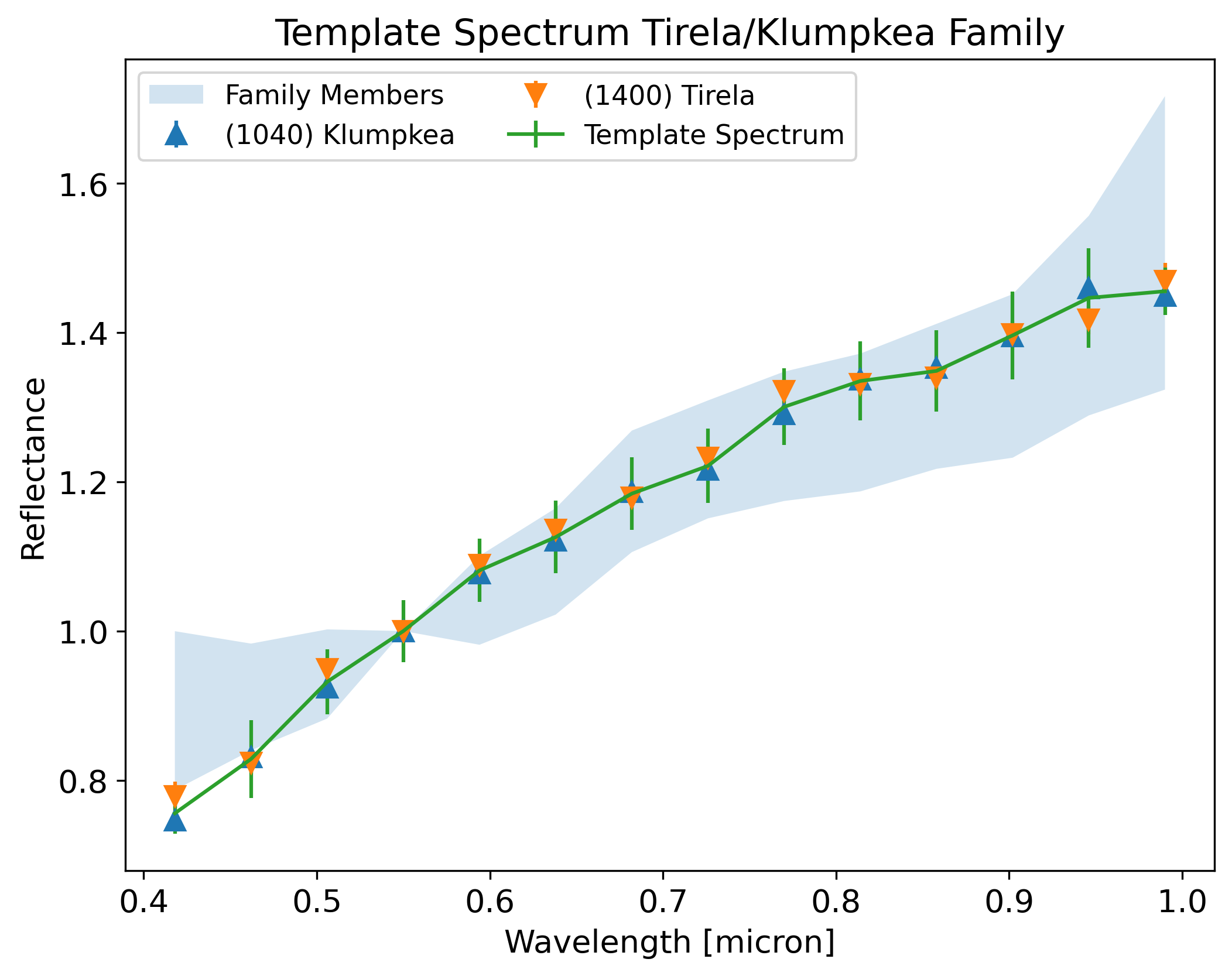}
      \caption{\small Template spectrum of the Tirela/Klumpkea family (in green). The reflectances have been computed by averaging the \gaia spectra of (1040) Klumpkea (in blue) and (1400) Tirela (in orange). The error bars represent the squared sum of their respective uncertainty. In light blue the region within which the spectra of the family members identified by our analysis lie within an interval of one standard deviation.
              }
         \label{TemplateTirela}
   \end{figure}

\begin{figure*}
   \centering
   \includegraphics[width=\textwidth]{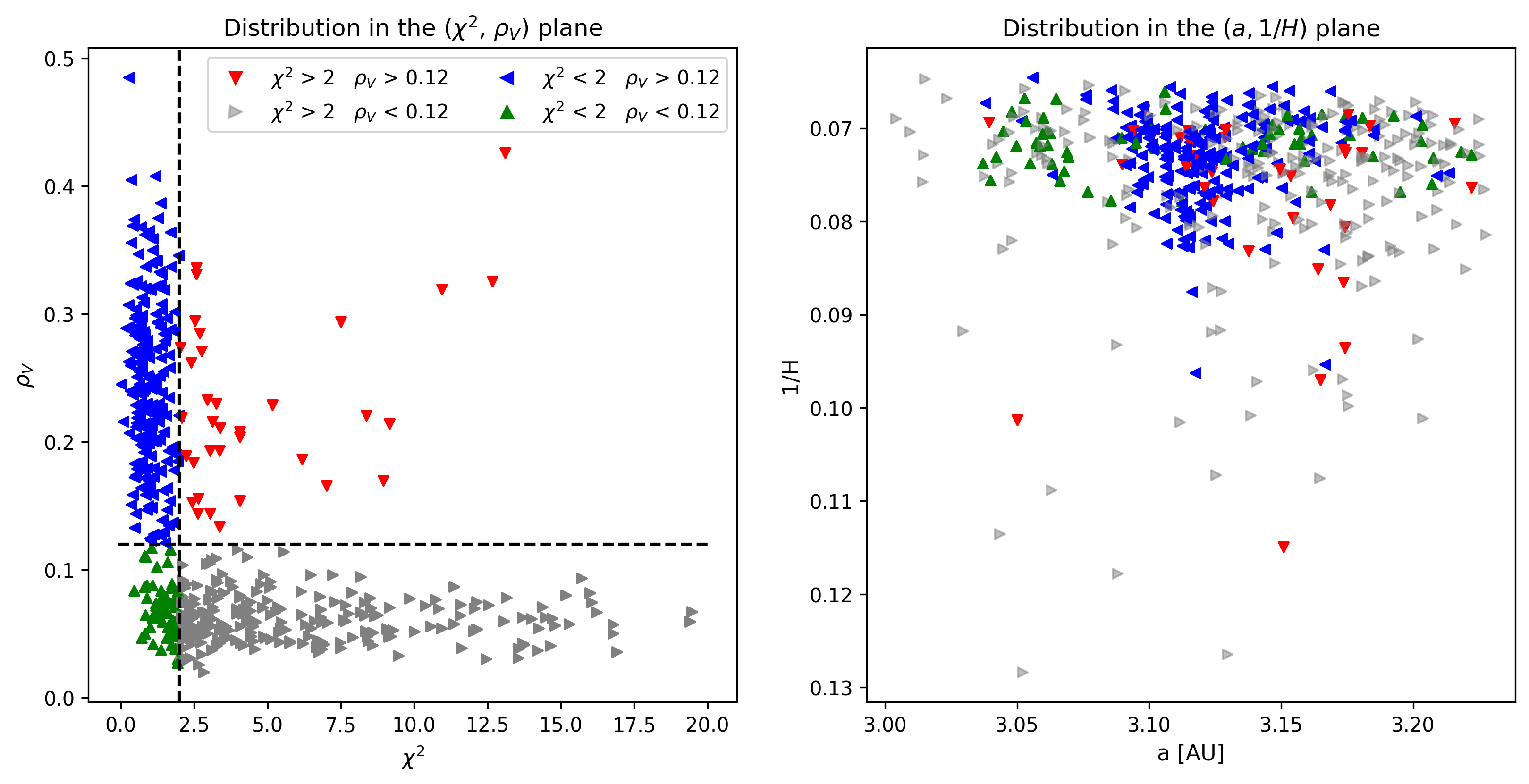}
      \caption{\small \newtext{$\chi^2$ distribution of the objects observed by \gaia within the Tirela/Klumpkea region.} Left panel: Distribution in the $(\chi^2, \rho_V)$ plane. Red points have $\rho_V$ > 0.12 and $\chi^2 > 2$, blue points have $\rho_V$ > 0.12 and $\chi^2 < 2$, green points have $\rho_V$ < 0.12 and $\chi^2 < 2$ and finally grey points have $\rho_V$ < 0.12 and $\chi^2 > 2$.
      Right panel: Distribution in the $(a, 1/H)$ plane of the points reported in the left panel.
              }
         \label{ChiSquareAlbedoTirela}
   \end{figure*}

\begin{figure}
   \centering
   \includegraphics[width=\hsize]{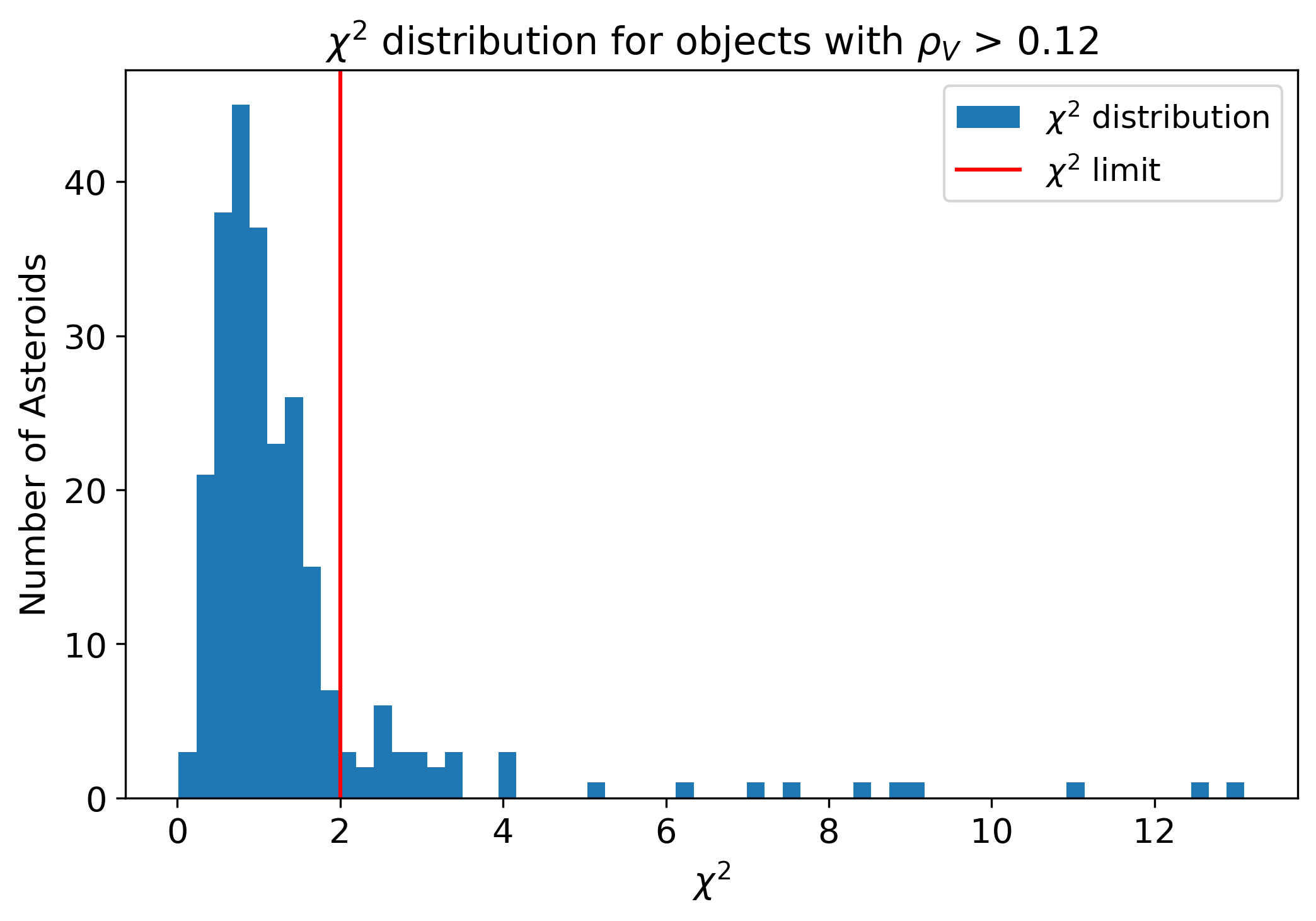}
      \caption{\small $\chi^2$ distribution for objects in the Tirela/Klumpkea region with $\rho_V > 0.12$. The red vertical line indicates the limit in $\chi^2$ to separate family members from the background.
              }
         \label{ChiSquareLimitTirela}
   \end{figure}

The reflectance spectrum from \gaia was segmented into 16 distinct wavelength bands, covering a range from 0.374 to 1.034 micrometers, each with a bandwidth of 0.044 micrometers. For every reflectance measurement, the dataset includes quality flags that detail the quality of the data and identify any encountered issues, in addition to providing information on flux uncertainties.
\\
To classify Solar System objects (SSOs), it is necessary to compare their reflectance spectra with established taxonomy templates. One method for classifying spectra is to use machine learning (ML) algorithms \citep{mahlke-2022}. Machine learning algorithms can learn from data and make predictions based on learned patterns. To train the model, labeled datasets are required, which are then used to classify new data. However, the task is complicated by systematic variations, particularly in the blue and red parts of the spectra \citep{tinautruano-2023}, as well as the presence of noise, and missing data values. These factors may lead to inaccurate classification of asteroids. \\
To address this issue, a generally adopted method is based on identifying and adjusting for systematic differences in \gaia spectra. This is done by comparing them with etalon spectra obtained, for example, from ground-based observations of the same asteroids. 
This method involves creating a compensation model, which may be affected by outliers and missing values. The model requires interpolations and extrapolations of \gaia reflectance data, which introduces additional limitations. Therefore, we employed an alternative strategy that does not rely on etalon spectra.

The method we used to prepare the training dataset was to utilize the \gaia spectra in their original form, which had been previously classified by ground-based spectral observations. These spectra were then used to train a random forest classifier \citep{breiman-2001}. The random forest classifier is an ensemble learning method that constructs multiple decision trees during training and outputs the class which is the mode of the classes of the individual trees. This classic ML classifier is well-suited for classifying \gaia spectra due to its robustness to overfitting and missed values, and its ability to handle large multilabel datasets with higher dimensionality.

The foundation of our asteroid taxonomy is built upon matching \gaia asteroids with the most comprehensive spectral database available to date, as compiled by \cite{mahlke-2022}. The spectral classification system used in their work was adopted, but consolidated into eight primary classes for analysis: A, C, D, K, L, S, V, and X types. This merging strategy streamlines classification following the \gaia spectra limits while preserving the essential distinctions between major asteroid types. This training dataset included more than 1,000 asteroids with known spectral types, 35 of which belong to the L- class.
The reflectance distribution template for L-type asteroids obtained from the \gaia dataset using this method is shown in Fig.~\ref{LTypeSpectra}.\newtext{The sudden change in the slope below 0.45 microns visible in the plot is a consequence of a flaw related to \gaia spectra, which mostly affects faint objects.}\\
The random forest classifier was trained on this dataset and the resulting model was used to classify the \gaia spectra. The identification of L-type asteroids in the visible domain presents some challenges, as their reflectance spectra in the visible can be sometimes similar to S-, K-, and D- types. Despite this difficulty, the accuracy of the classification of L-type asteroids with our approach reaches 87\%, which is a good result given the complexity of the task. The classifier outputted the two most probable classes for each asteroid, along with the associated probabilities. The classification results were then used to identify the most probable spectral type for each asteroid, which was then used to determine the asteroid's membership in the L-type family. \\
More details on the classification method can be found in 
Sergeyev \& Carry (in preparation).

\section{Application to the Klumpkea/Tirela collisional family} \label{SectionTirela}

\subsection{Membership}

Tirela is an asteroid family first identified by \cite{nesvorny-2005} and successively reidentified in the space of proper elements by different authors, such as \cite{masiero-2013} and \cite{milani-2014}, who used different implementations of the HCM algorithm. The latter ones in particular found a different membership to the family and renamed it the Klumpkea family. \\
The Tirela/Klumpkea family is located at the edge of the outer Main Belt, at semi-major axis $a=3.12$ AU, and it possesses high eccentricity and inclination, $e=0.20$ and $i=16.8^{\circ} $. The family is characterized by a high geometric albedo (0.2 - 0.3), whereas the background asteroids in the same region generally have low albedos and belong to the C- complex. The spectra of Tirela/Klumpkea members are reported in \cite{mothediniznesvorny-2008} and in \cite{devogele-2018}, who report the family to consist of L- types. \cite{milani-2016}, using the V-shape method described in \cite{spoto-2015} determined an age of about 660 Myr.
   
To identify the Tirela/Klumpkea family using \gaia spectra alone, we first considered all asteroids contained within a region in the 3D space of proper elements centered on the cluster of identified Tirela/Klumpkea members with 3.00  AU $< a <$ 3.23  AU, $0.15 < e < 0.25$ and $0.27 < \sin(i) < 0.33$. In addition to the family members, this wide region contains numerous background asteroids and other families (or at least part of them). The proper elements $a$, $e$, and $\sin(i)$ and the absolute magnitudes, $H$, were retrieved from the Asteroid Family Portal (AFP, \citealt{novakovic-2022}), while the albedos, $\rho_V$, were extracted from the WISE/NEOWISE survey \citep{masiero-2011}. For objects observed more than once by WISE, the weighted average of their measurements was taken.\\
\newtext{Overall,} 8175 objects were found to be located within the selected region, 1574 of which were listed as part of the family in the Asteroids-Dynamic Site (AstDyS, \citealt{milani-2014}). Among the total sample, 704 objects presented a spectrum in the \gaia DR3, while only 41 objects had a spectrum published from other sources. \newtext{A total of} 1887 objects instead already had colors determined from photometric observations.

\newtext{Among the total sample, 1928 objects possessed a NEOWISE albedo. Similarly, out of the 704 objects in the \gaia sample, 548 had a NEOWISE albedo.} The distribution of the albedos is reported in the left panel of Fig.~\ref{TirelaAlbedo}, where the complete sample is shown in blue, the set having \gaia spectra in yellow and the family members included in AstDyS in red. A bimodal distribution, expected in the selected volume of proper elements, stands out, with a separation around $\rho_V = 0.12$. Given the position in the outer belt, the narrow peak at low albedos corresponds to dark background asteroids. The broad peak at higher albedos consists of Tirela/Klumpkea family members plus a few asteroids belonging to the S- complex; given the rough spectral similarity with the L- types, it could also be family members missed by the HCM classifiers. AstDyS's membership contains few low albedo objects, which are likely interlopers.

The left panel of Fig. \ref{TirelaAlbedo} also reports in green the distribution of the family members included in the AFP classification system. This scheme, also based on an HCM algorithm, includes a large number of low albedo objects, which very likely are interlopers. We therefore discarded the AFP classification system and retained AstDyS as the only literature source of comparison with our work. Throughout the paper we will use the abbreviations Mi16 when specifically referring to the paper \cite{milani-2016} and HCM when referring to the classification scheme first developed in \cite{milani-2014} and implemented in AstDyS. Moreover,  to update the membership with more recent proper elements (extracted from AFP, version of October 2022) and to check the detailed hierarchy of the clustering process, we also re-ran the HCM independently. 

The correlation between high albedos and family membership is confirmed by the distribution in the proper elements space, as shown for the $(a, e)$ plane in the right panel of Fig. \ref{TirelaAlbedo}, where the points have been subdivided according to the albedo. Objects with $\rho_V > 0.12$ concentrate in the family core centered between 3.09 AU and 3.15 AU. Some scattered high albedo objects are also present, in a range of semi-major axes similar to the family core, and in the resonances. Objects with $\rho_V < 0.12$ instead are scattered without a preferred distribution, except for a small cluster at 3.05 AU which is the Aegle family. This analysis proves that the family members of Tirela/Klumpkea can be separated from the background using the albedo, which in turn correlates with the spectral type. 

This evidence motivated us to develop a procedure to identify the family members using \gaia spectra alone, without the support of albedos. This allowed us, in principle, to find family members whose albedo has not yet been measured. The procedure is based on two selection filters, which an object must pass in order to be considered a family member. The first filter quantifies the spectral similarity between an object and the largest members of the family, whose spectra are combined to create a family template spectrum.
The second filter is instead based on the color taxonomy described in Sect. \ref{Classyfing Gaia Spectra}. Only objects whose spectral types are the same as the largest fragments of the Tirela/Klumpkea family are considered to be part of it.

\begin{figure}
   \centering
   \includegraphics[width=\hsize]{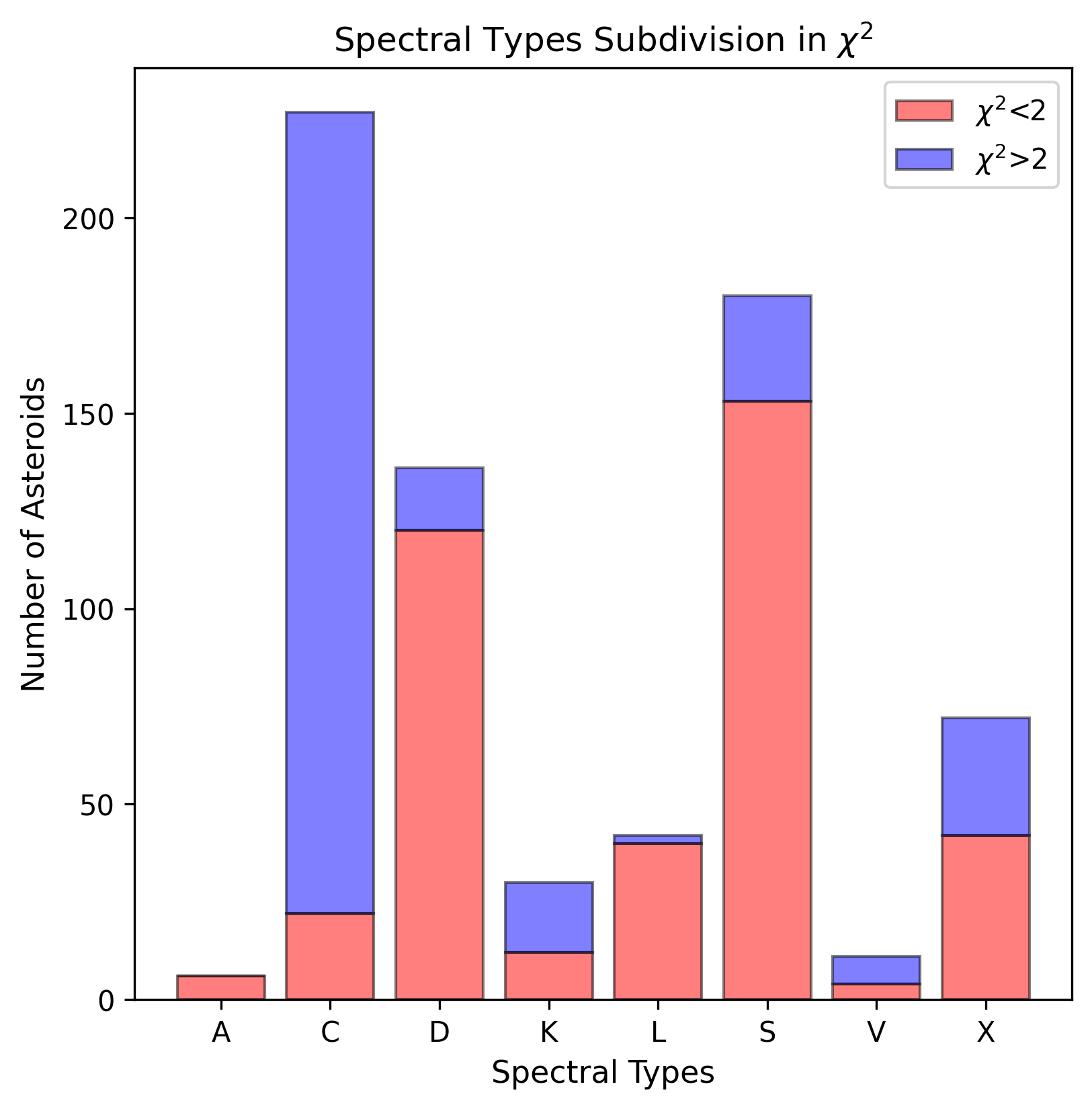}
      \caption{\small Spectral types of the objects observed by \gaia in the Tirela/Klumpkea region. In blue, we show objects with $\chi^2 > 2$; in red objects, with $\chi^2 < 2$.
              }
         \label{TaxonomyTirela}
   \end{figure}

\begin{figure*}[ht]
   \centering
   \includegraphics[width=\textwidth]{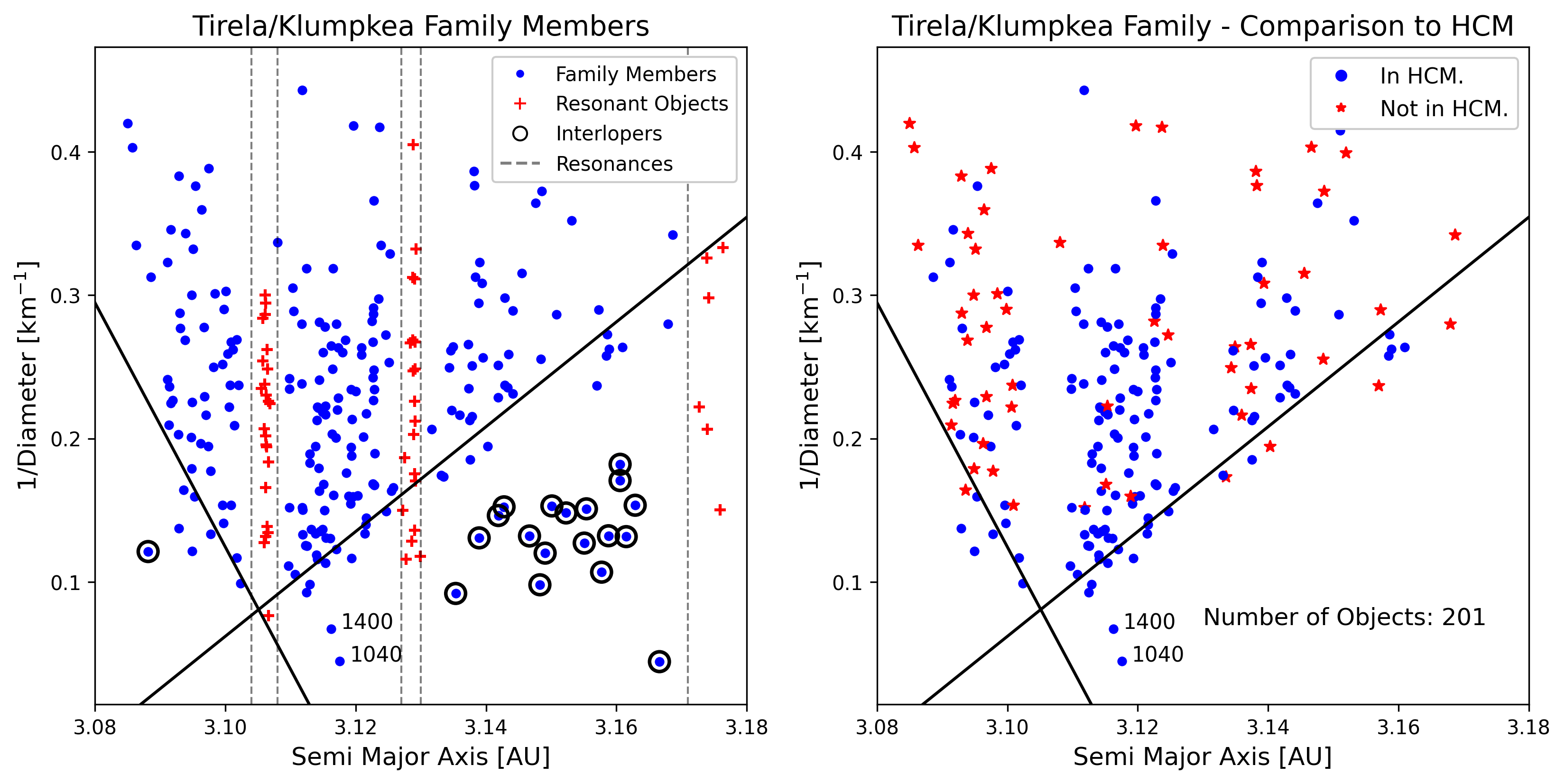}
      \caption{\small \newtext{V-shape of the Tirela/Klumpkea family members retrieved from this work.}
      Left panel: Distribution in the $(a, 1/D)$ plane of the family members identified from \gaia spectra (in blue). Red crosses are resonant objects, while objects circled in black on the outer side are excluded from the family for their distance from the V-shape. The only excluded object on the left is rejected by the V-shape fit procedure. The two black lines are the fitted V-shape for $N=13$ (see Fig. \ref{FitVShapeTirela}). \newtext{The numbers mark the positions of (1040) Klumpkea and (1400) Tirela.}
      Right panel: Distribution in the $(a, 1/D)$ plane of Tirela/Klumpkea members retrieved from \gaia observations. Blue points are objects in common between this work and the HCM, while red stars are objects that are not linked by the HCM.
              }
         \label{VShapeTirela}
   \end{figure*}
   
The family template spectrum is shown in Fig.~\ref{TemplateTirela}, whose reflectance has been computed from a weighted average of the spectra of (1400) Tirela and (1040) Klumpkea, the two largest fragments of the family,  shown in orange and in blue, respectively. \cite{popescu-2012} propose five possible statistical criteria for curve matching based on the $\chi^2$ method and the correlation coefficient. We tested all these methods obtaining very similar results and we chose to use \newtext{a modified version of} the classical $\chi^2$, which was the one with the best performance. We stress here that an optimal combination of multiple criteria might perform even better, however it would probably affect the results only very marginally. Such an optimization is beyond the scope of this article and may be implemented for future studies. In addition, this is just a first step to filter out interlopers from the family. The important selection to identify family members is in fact conducted using the color taxonomy.
\newtext{The modified version of the classical $\chi^2$ that we defined is based on the procedure adopted in \citet{avdellidou-2022}}. For each object $\chi^2$ is computed as:

\begin{equation}
    \chi^2 = \sum_{\lambda} \frac{[G(\lambda)-\beta T(\lambda)]^2}{\sigma_{G}(\lambda)^2+\sigma_T(\lambda)^2} \ \  \ \textrm{with} \ \ \ \beta = \frac{\sum_{\lambda}T(\lambda)G(\lambda)}{\sum_{\lambda} T(\lambda)^2}
    \label{EqChiSquare}
,\end{equation}

where $G(\lambda)$ is the \gaia reflectance, $\sigma_G$ its uncertainty, $T(\lambda)$ is the family template reflectance, and $\sigma_T(\lambda)$ its uncertainty; also, $\beta$ is a factor taking into consideration that the spectra could be arbitrarily normalized. \newtext{However, this is not the case here since both \gaia spectra and the template are normalized at 550 nm and therefore $\beta = 1$.}

Family members in principle should present spectra similar to the template and therefore should have small $\chi^2$ values. On the other side, background objects should differ from the template, resulting in large $\chi^2$ values. The left panel of Fig.~\ref{ChiSquareAlbedoTirela} reports the distribution in the $(\chi^2, \rho_V)$ plane of the objects observed by \gaia with a NEOWISE albedo. The right panel reports the same points in the $(a, 1/H)$ plane where the V-shape of the family should stand out. The $(\chi^2, \rho_V)$ plot clearly shows two main clusters, again separated in two albedo categories, plus some scattered data points.\\
We further divide the distribution into four subgroups, by exploiting both the albedo bi-modality (split at $\rho_V = 0.12$) and a threshold on $\chi^2$. This threshold, $\chi^2 = 2$, has been chosen from the distribution of $\chi^2$ for objects with albedo $\rho_V > 0.12$, reported in Fig. \ref{ChiSquareLimitTirela} and clearly showing a large number of asteroids at low $\chi^2$, plus a queue of outliers. Of course, the exact value needed to discriminate spectra similarity is arbitrary and can be ambiguous for objects that are very close to it.\\
These criteria result in the following partitioning:

\begin{itemize}
    \item Grey points in Fig.~\ref{ChiSquareAlbedoTirela} (left panel): low albedo and large $\chi^2$. They are the background asteroids belonging to the C- complex uniformly distributed in space.
    \item Red points: high albedos and large $\chi^2$. They are S- complex asteroids with spectra deviating from the family template.
    \item Blue points: high albedos and low $\chi^2$. They belong to the S- complex and they present spectra similar to the family template. In the right panel, they are clearly compatible with a V-shape for the family.
    \item Green points: low albedos and low $\chi^2$. This may appear contradictory, suggesting that dark background asteroids exhibit spectra similar to the family template. However, in reality, they are faint objects with spectra affected by large error bars, thus resulting in a low value of $\chi^2$.
\end{itemize}

\begin{figure*}
   \centering
   \includegraphics[width=0.93\textwidth]{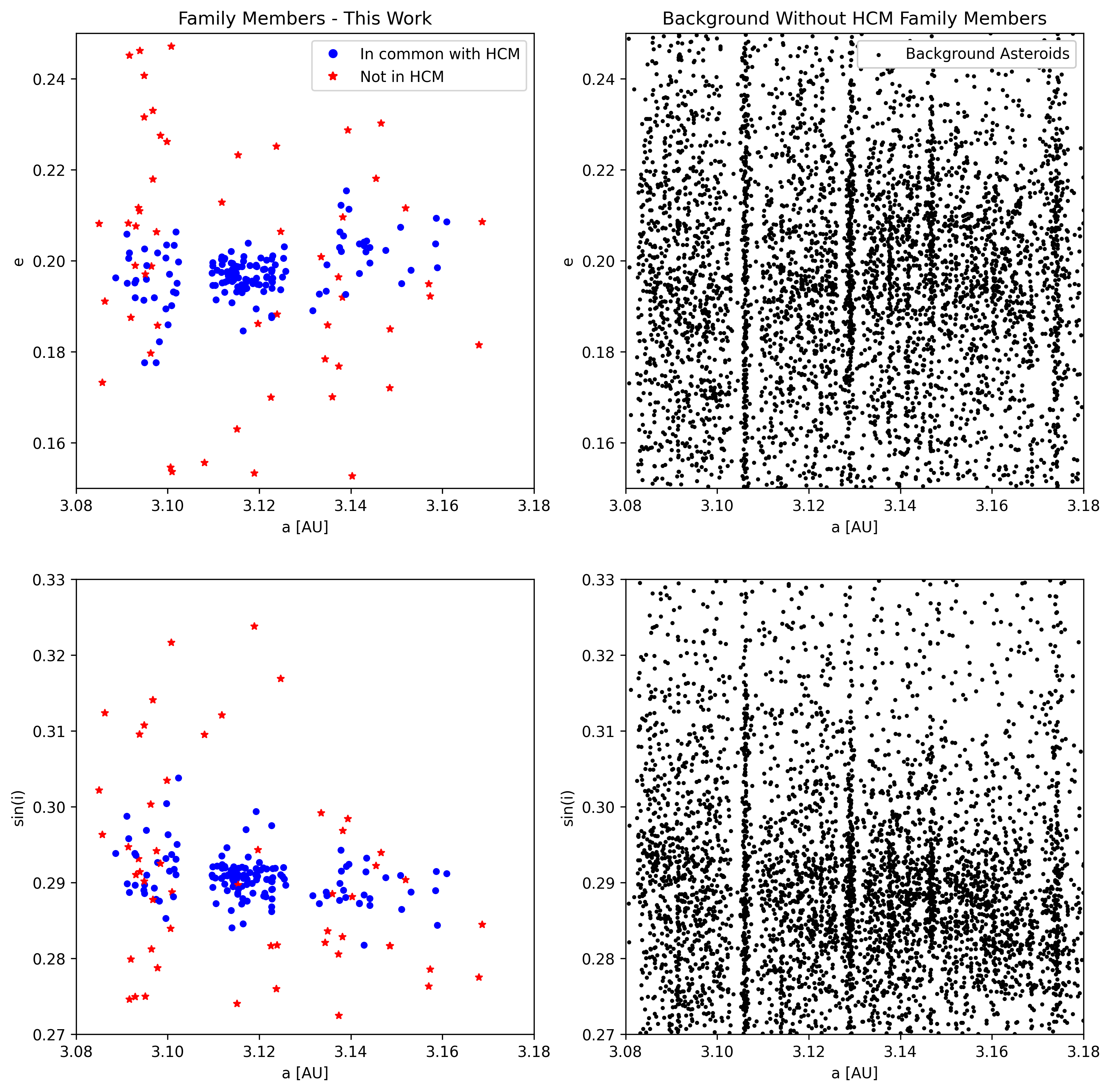}
      \caption{\small Distribution in the $(a, e)$ and $(a, \sin(i))$ planes for objects located in the Tirela/Klumpkea region. The left column reports the distribution of the family members retrieved from this work. Blue points are in common with the HCM, red objects are not. The right column reports the distribution of all asteroids located within the Tirela/Klumpkea region from which the HCM family members have been removed.
              }
         \label{HaloTirela}
   \end{figure*}

\begin{figure}
   \centering
   \includegraphics[width=\hsize]{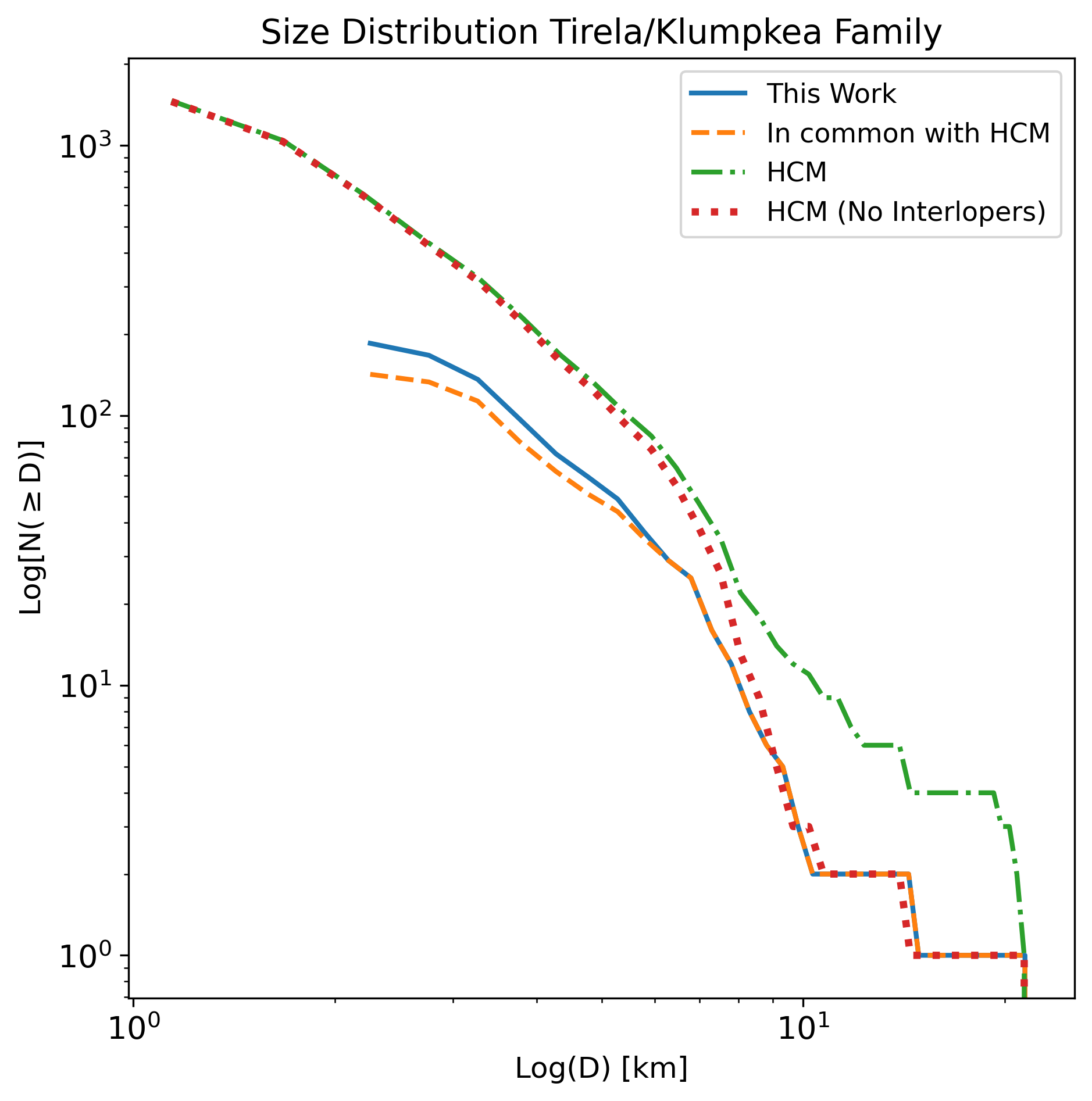}
      \caption{\small Size distribution for the family members retrieved from this work (blue line), for the family members in common in this work and the HCM (orange dashed line), for all the objects identified by the HCM (green dash-dotted line) and for the objects identified by the HCM without the interlopers of Table \ref{table:MilaniInterlopers} (red dotted line).
              }
         \label{SizeDistributionTirela}
   \end{figure}

\begin{figure}
   \centering
   \includegraphics[width=\hsize]{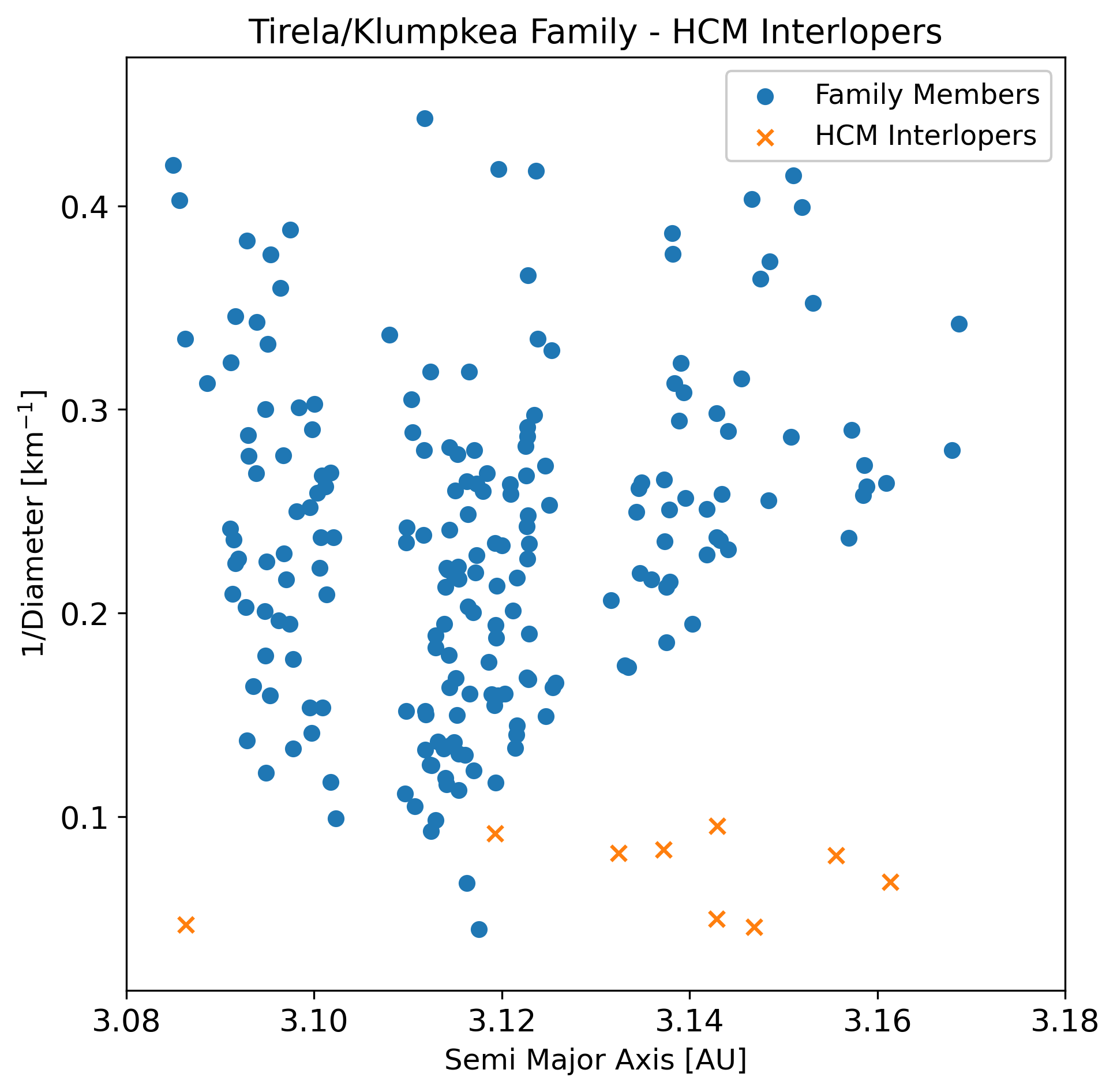}
      \caption{\small Distribution in the $(a, 1/D)$ plane of the HCM interlopers (orange crosses) and the Tirela members identified in this work (blue points).
              }
         \label{MilaniInterlopersTirela}
   \end{figure}

From this distribution, it is clear that a selection based only on spectral similarity would include both low- and high-albedo objects (the last two categories above). To address this aspect, we conducted an additional selection based on the taxonomy derived from the spectra, outlined in Sect.~\ref{Classyfing Gaia Spectra}.\\
The resulting distribution of spectral types is shown by Fig.~\ref{TaxonomyTirela}, divided into two categories by the $\chi^2 > 2$ threshold. A good correlation between the $\chi^2$ value and the spectral types is observed, with S- complex objects exhibiting spectra similar to the family template ($\chi^2 <2$) and C- complex objects with spectra diverging from the template ($\chi^2 >2$). There are a few C- and X- types with $\chi^2 < 2$: these are the green points in Fig. \ref{ChiSquareAlbedoTirela}, faint objects with noisy spectra. In fact, they present an average absolute magnitude of $H = 14.0$, slightly less than the average of the total sample, $H = 13.5$. In addition, the average S/N of their spectra is 19.6, notably smaller than the average of the other points, $S/N = 46.8$, and close to the limit of $S/N = 13$ defined by \cite{Gaiacollaboration-2023} for the publication of spectra in \gaia DR3. The distribution also shows that L- type asteroids degenerate with S- types and partially with D- types, probably due to the large dispersion of the spectra reflectances in the near-infrared \citep{Gaiacollaboration-2023}.\\
To understand the degree of degeneracy between the L- and S- types and the other classes we analyzed the second most probable class retrieved from the classifier for the objects with $\chi^2 <2$ and initially classified as A-, C-, D-, K-, V-, and X- types. We obtained that about 2/3 of D- types, 2/3 of K- types, half of X- types, and all A- types received as second classification the L- or the S- classes. The C- and V- types instead turned out to present spectra with a shape very different from the family template.

\begin{figure*}
   \centering
   \includegraphics[width=0.95\textwidth]{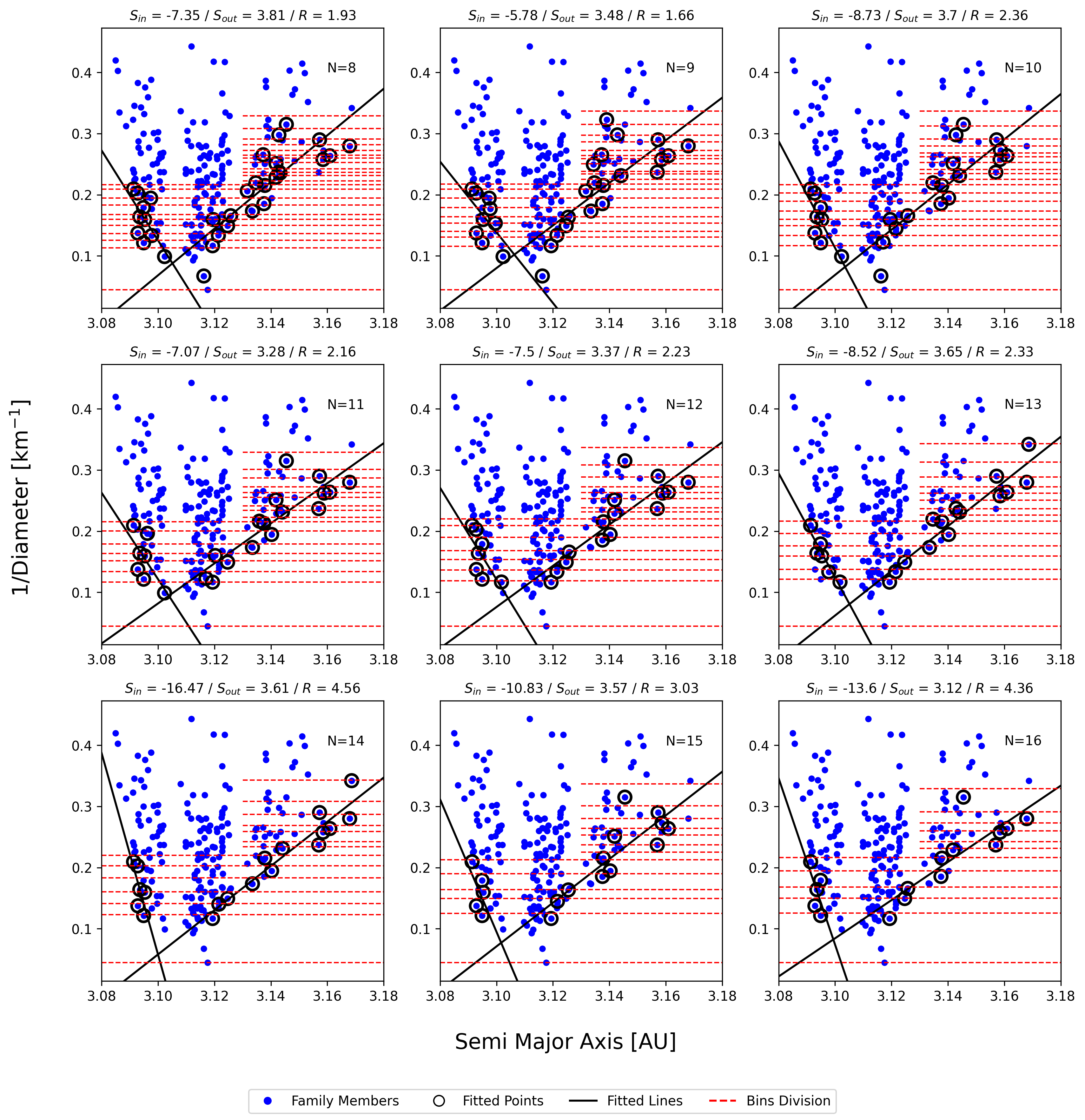}
      \caption{\small Illustration of the fitting procedure of the V-shape of the Tirela/Klumpkea family for $N$ ranging between 8 and 16, where $N$ is the number of objects contained within a bin. The blue points are the family members, the black circles are the points used for the fit, the black lines are the fitted V-shapes and the red dashed lines are the bin subdivisions. The title of each panel reports the values of the inner slope, the outer slope, and their ratio.
    }
         \label{FitVShapeTirela}
   \end{figure*}

We therefore decided to consider as family members the S- and L- types with $\chi^2 < 2$ and among the A-, D-, K-, and X- types with $\chi^2 <2$ those which received L- or S- as the second most probable class. This allows to include in the family L- types that are misclassified by the taxonomy or that present flawed \gaia spectra, accepting the minor risk of including in the family faint interlopers which may present a small $\chi^2$ only because of highly noisy spectra. All C- and V- types have instead been discarded from the family.

The objects selected in this way are clustered between 3.08 AU and 3.18 AU, which have been adopted as boundaries in semi-major axis for the family. \newtext{This region is cut by numerous high-order mean motion resonances with Jupiter, the most important of which are 13/6J, 15/7J, and 21/10J \citep{radovic-2017}. We thus defined three regions at 3.104 AU $< a <$ 3.108 AU, 3.127 AU $< a <$ 3.130 AU, and 3.17 AU $< a <$ 3.18 AU within which an asteroid is considered to be in resonance.} Resonant objects may in principle belong to the family, but their eccentricities have been pumped up (or down) and their evolution in semi-major axis, due to the Yarkovsky drift, has been halted \citep{nesvorny-2002}. For the analysis that follows, objects in resonances are neglected, regardless of their spectral types.

To identify a V-shape for the family in the $(a, 1/D)$ plane, we computed the diameters $D$ of the selected objects from their albedo $\rho_V$. The albedo is either extracted from NEOWISE, when available, or assigned from the average albedo of the family, $\rho_V = 0.23 \pm 0.08$. This value is also compatible with $\rho_V = 0.204 \pm 0.100$ reported in Mi16. 
The resulting distribution in the $(a, 1/D)$ plane of the family members is reported in the left panel of Fig.~\ref{VShapeTirela}. A rather well-separated cluster of large asteroids on the outer side, circled in black, respect the two criteria on spectra similarity and therefore are selected as family members by our procedure. However, they are well separated by a gap from the V-shape of the family, and moreover are almost as big as (1400) Tirela and (1040) Klumpkea, which would imply a very improbable size distribution. These objects were also not linked to the family by the HCM. We therefore exclude these objects from the family. The V-shape of the family is quite well defined in the outer side until small diameters, while in the inner side it is cut by the 11:5 mean motion resonance with Jupiter. Interestingly, the two largest fragments of the family are slightly shifted in semi-major axis compared to the vertex of the V-shape.\\
The blue points in the left panel of Fig. \ref{VShapeTirela} are the final family members retrieved from \gaia observations, which are used in a later step to analyze to place constraints on the age of the family. They are reported in the right panel of Fig. \ref{VShapeTirela}, where they are compared to the family identified by the HCM. It is clear that at diameters $D < 7$ km, our method identifies objects that the HCM does not link to the family, but that are well compatible with the V-shape. We detected 173 objects within the family for $D < 7$ km, 54 of which, the 31\%, were not identified by the HCM. For $D < 3$ km, we instead identified 24 objects, 14 of which were not linked by the HCM.\\
A total of 201 family members were retrieved from the \gaia spectra and the color taxonomy. The complete family membership for the Tirela/Klumpkea family is reported in the Zenodo repository \footnote{\url{https://doi.org/10.5281/zenodo.12704069}}.

The spatial distribution of these objects in the $(a, e)$ and $(a, \sin(i))$ planes is reported in the left column of Fig. \ref{HaloTirela}, where the colors and symbols follow the right panel of Fig.~\ref{VShapeTirela}. The objects in common between our work and the HCM form the family's core, while the objects we retrieved are more dispersed and harder to identify for the HCM. The right column of Fig. ~\ref{HaloTirela} includes all asteroids in the proper element volume around the family. Although objects classified as family members by the HCM have been subtracted, a diffuse halo extending the family volume is visible.
The distribution of our additional family members appears qualitatively compatible with this diffuse halo. 

\begin{table}
\caption{\newtext{Interlopers included in the HCM family.}}             
\label{table:MilaniInterlopers}      
\centering                          
\begin{tabular}{c c c}        
\hline\hline                 
\small{Number} & \small{Spectral Type} & \small{NEOWISE Albedo} \\    
\hline                       
    (3667) & D & 0.06 \\     
    (7411) & C & 0.04 \\
    (18399) & X & 0.04 \\
    (24878) & C & 0.06 \\
    (29391) & C & 0.05 \\ 
    (42521) & C & 0.05 \\ 
    (44519) & C & 0.04 \\ 
    (65676) & X & 0.06 \\ 
    (160278) & C & 0.04 \\ 
\hline                                   
\end{tabular}
 \begin{tablenotes}
        \footnotesize
        \vspace{3 px}
        \item \newtext{Interlopers larger than 10 km included in the family by the HCM. The columns report the number of the asteroid, the spectral type retrieved from the color taxonomy and the NEOWISE albedo.}
    \end{tablenotes}
\end{table}

We also compared the new cumulative size distribution for the membership that we establish to the one by the HCM, reported in Fig. \ref{SizeDistributionTirela}. As expected, the additional members identified at small diameters by our selection produce a modest slope increase at $D <$ 7~km due to the halo objects.\\
The HCM distribution instead includes more objects and extends to smaller diameters, since the HCM is less limited in absolute magnitude than \gaia observations. However, the most interesting aspect is that the HCM includes in the family at least eight objects as large as (1400) Tirela and (1040) Klumpkea, which were instead discarded by our selection. As listed in Table \ref{table:MilaniInterlopers}, they belong to the C- complex and present a low albedo. They are thus very likely to be interlopers that can be identified as such only from their spectral properties. This conclusion is supported by their position relative to the V-shape of the family (Fig.~\ref{MilaniInterlopersTirela}), which is generally external to the family boundaries and far from the vertex of the distribution. The resulting size distribution without these interlopers is much closer to what is observed in other families, with only one or two dominating remnants. 

\newtext{Among these eight interlopers, six were already identified by \cite{radovic-2017} using an automated procedure to exclude interlopers from asteroid families. However, \cite{radovic-2017} also report (1400) Tirela to be an interloper, which we included in the family instead. This is coherent with the spectra of few Tirela/Klumpkea members reported in \cite{devogele-2018}, which underline the spectral similarity between (1400) Tirela and the other family members.}\\
We also stress here that other interlopers could remain in the HCM data set, but it is more difficult to identify them for D$<$7--10~km, due to the lower quality of the spectra and the absence of albedo measurements. \\
The family membership retrieved by our run of the HCM is very similar to the one of AstDyS and, therefore, the overall results do not change between the two versions.

\subsection{Age determination}

In order to fit the V-shape and constrain the age of the Tirela/Klumpkea family, we developed a procedure based on the work by \cite{milani-2014} and \cite{spoto-2015} to determine the slopes of the sides of the V-shape. The first step requires us to select the intervals of semi-major axis and size defining the V-shape.
The boundaries in the semi-major axis are taken between 3.08 AU and 3.18 AU, the same as in Mi16, while the upper limit in $1/D$ is taken where the sides depart from the  V-shape and become vertical. Given the limited number of members within the family, the choice of the $1/D$ significantly limits the reliability of the boundary slope, especially for the inner side, which is clearly less extended and defined. This leads us to adopt $1/D = 0.22 \ $km$^{-1}$ for the inner side and $1/D = 0.35 \ $km$^{-1}$ for the outer side. The y-axis is then divided into $n$ bins, $i=\left(1,2,...,n \right) $, such that each bin contains the same number of objects $N$. \cite{spoto-2015} implemented an automatic procedure which, given the number of asteroids inside the family, provides the best value of $N$ for the subdivision in bins. In this work, we explored a range of $N$ instead, repeating the binning for $N$ between 8 and 16. 

Inside each bin, the points characterized by the smallest (inner side) and largest (outer side) semi-major axes are selected. Apart from small $N$, the two largest bodies (Tirela and Klumpkea) are never selected for the computation of the slopes. Automatically including them in the fitting process could be an option, however, this would give excessive weight to the few largest objects, introducing a clear bias due to their displacement with respect to the vertex. This shift could be related to the violent fragmentation event with a non-negligible ejection velocity of the largest fragments \citep{spoto-2015}.

\begin{figure}
   \centering
   \includegraphics[width=\hsize]{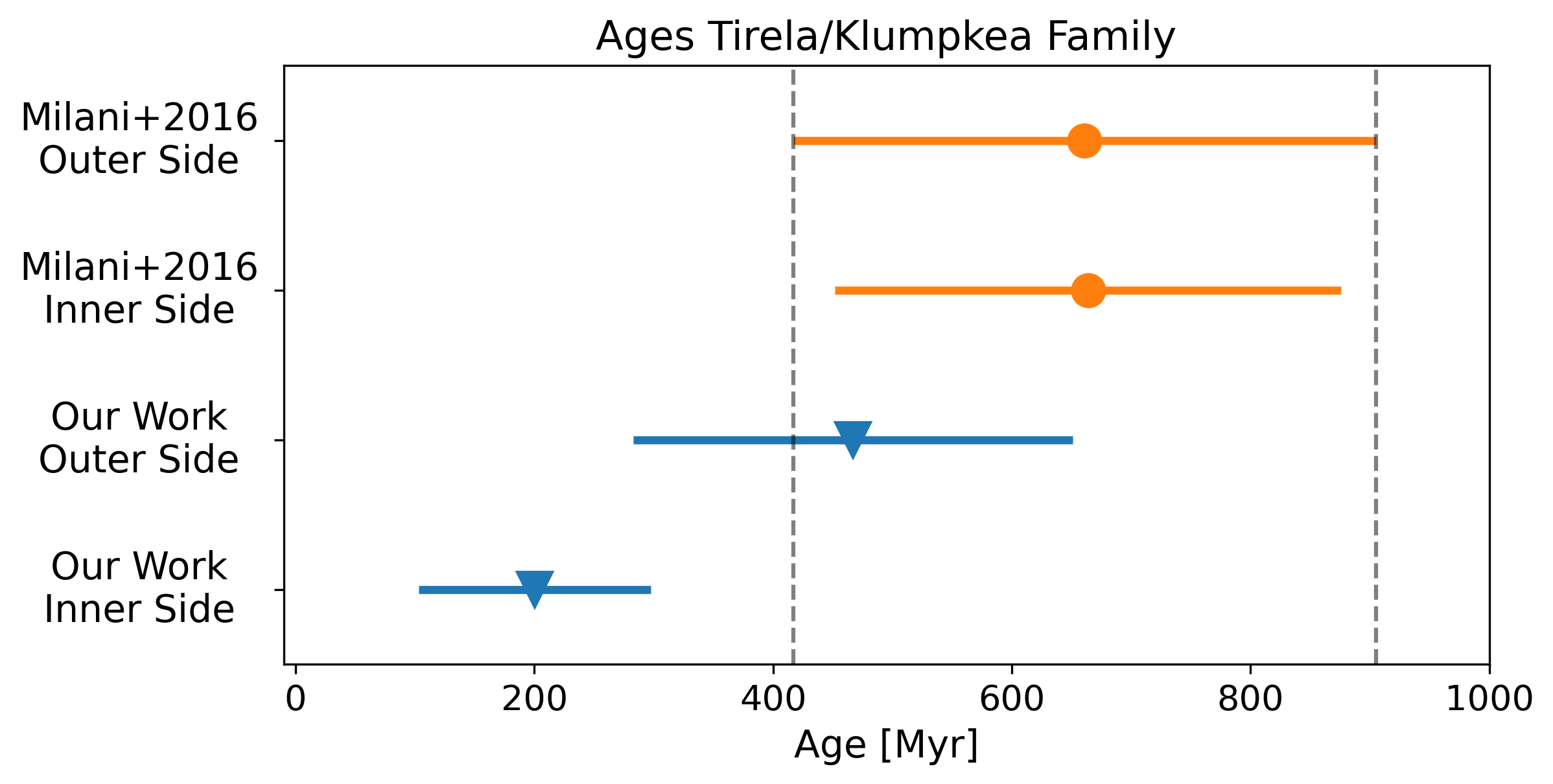}
      \caption{\small \newtext{Ages of the Tirela/Klumpkea family determined in this work, in blue, compared to the ages reported in Mi16, in orange. The black dashed vertical lines report the uncertainty on the age reported in Mi16.}
              }
         \label{AgesTirela}
   \end{figure}

\begin{figure}
   \centering
   \includegraphics[width=\hsize]{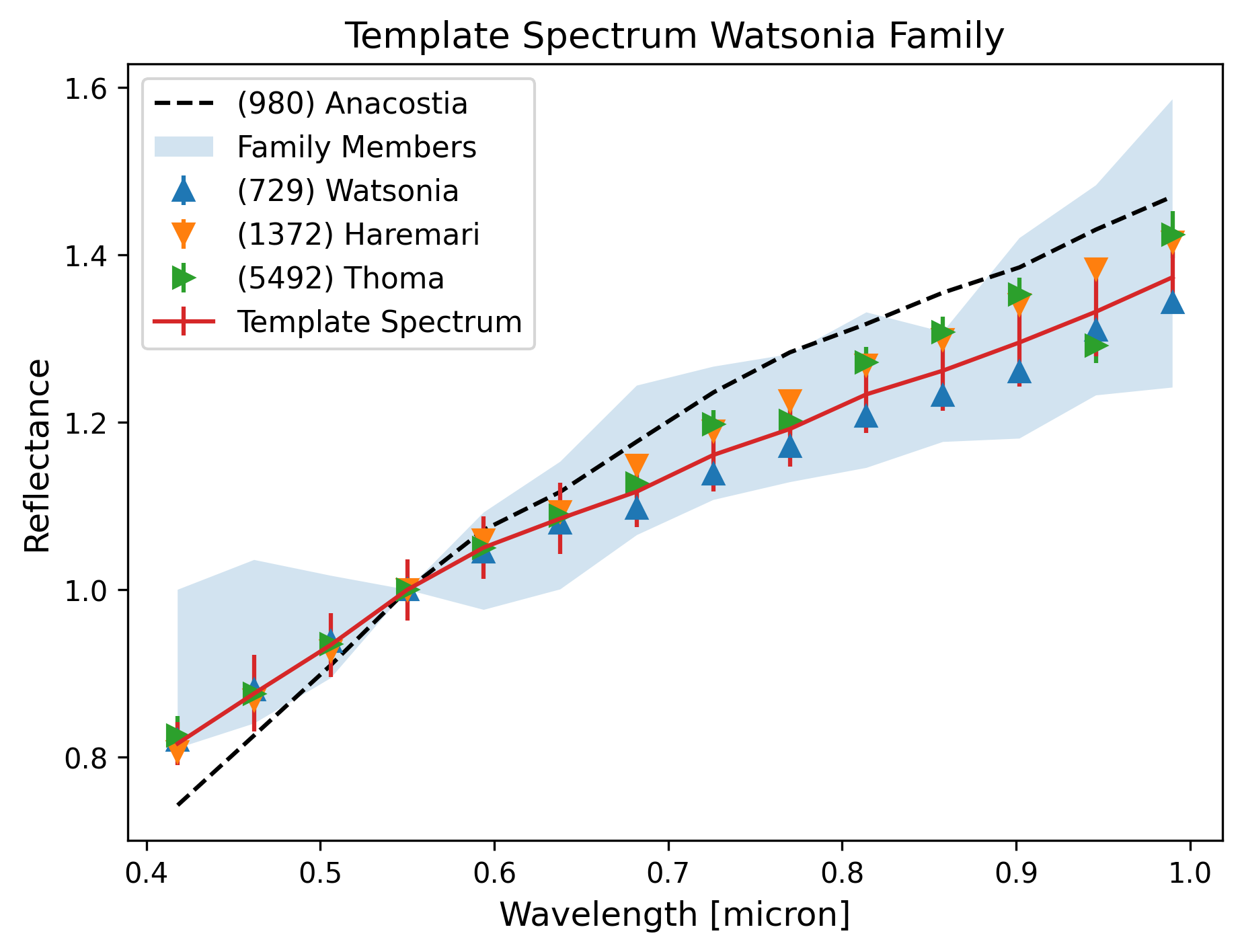}
      \caption{\small Template spectrum of the Watsonia family (in red). The reflectances have been computed by averaging the \gaia spectra of (729) Watsonia (in blue), (1372) Haremari (in orange), and (5492) Thoma (in green). In light blue, the region within which the spectra of the family members identified by our analysis lie within an interval of one standard deviation.
      The black dashed line reports the spectrum of (980) Anacostia.
              }
         \label{TemplateWatsonia}
   \end{figure}

\begin{figure*}
   \centering
   \includegraphics[width=\textwidth]{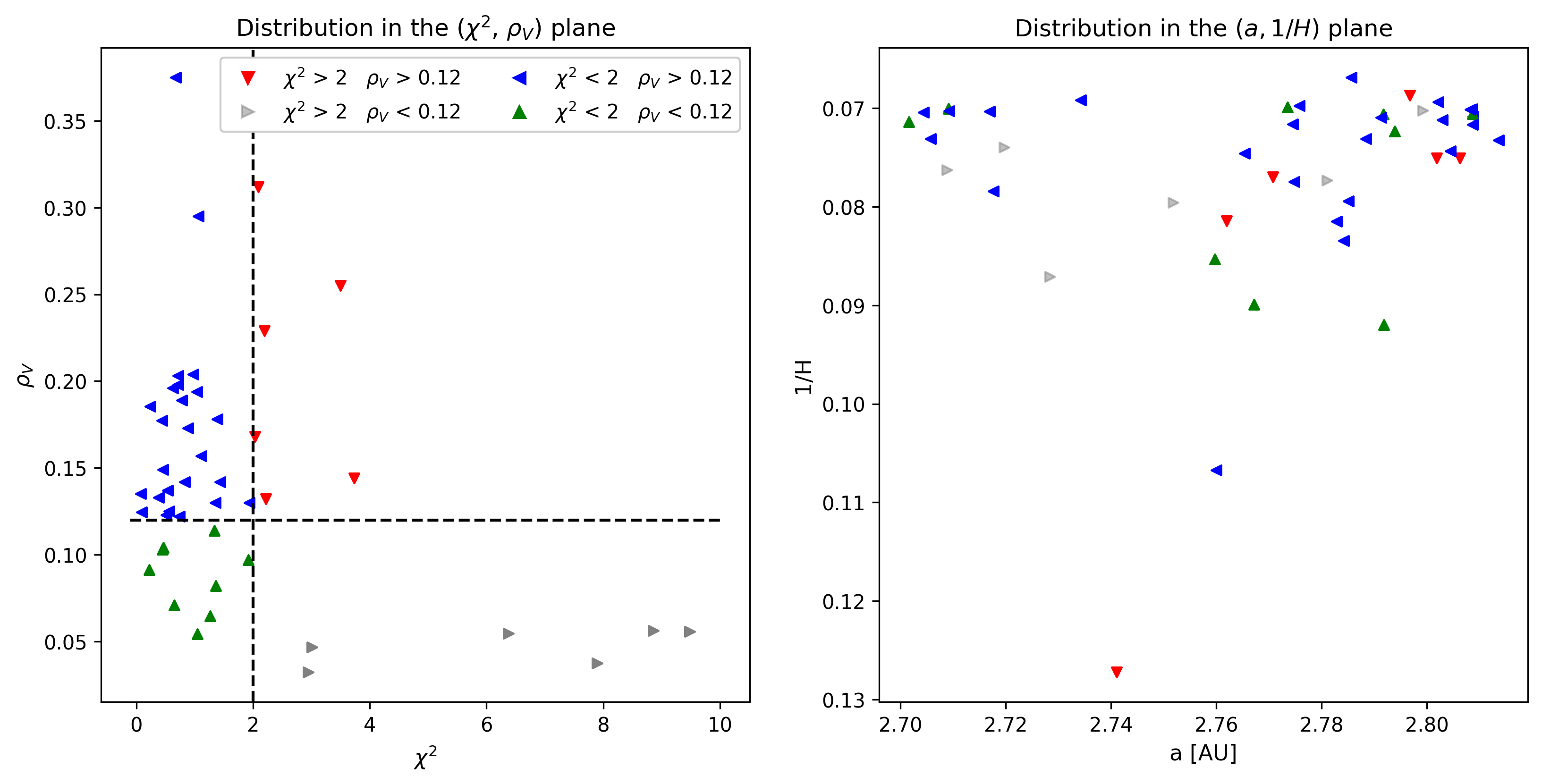}
      \caption{\small \newtext{$\chi^2$ distribution of the objects observed by \gaia within the Watsonia region.} Left panel: Distribution in the $(\chi^2, \rho_V)$ plane. Red points have $\rho_V$ > 0.12 and $\chi^2 > 2$, blue points have $\rho_V$ > 0.12 and $\chi^2 < 2$, green points have $\rho_V$ < 0.12 and $\chi^2 < 2$ and finally grey points have $\rho_V$ < 0.12 and $\chi^2 > 2$.
      Right panel: Distribution in the $(a, 1/H)$ plane of the points reported in the left panel.
              }
         \label{ChiSquareAlbedoWatsonia}
   \end{figure*}

The two sets of points are then fitted through a linear regression algorithm, which finds the best straight lines approximating the two sides of the V-shape. For each side, we compute the residuals of the linear regression and their standard deviation, $\sigma_r$. Only one outlier is found at the level of $2\,\sigma_r$, the threshold chosen for rejection, the point circled in black on the inner side in the left panel of Fig. \ref{VShapeTirela}. This point (\newtext{which is not included in the family by the HCM either}) is therefore discarded and the linear regression is repeated on the new set of data points. The resulting V-shapes are indicated in black in Fig. \ref{FitVShapeTirela}, which summarises the whole procedure and reports the slope of the inner side, $S_{in}$, the slope of the outer side, $S_{out}$, and their ratio, $R$.\\
The slope of the outer side remains almost constant with varying $N$, with only small fluctuations that can be explained by the different points used for the fit in each case. The inner side is instead less self-consistent since the fit is conducted over a much smaller range in $D$. The inner slope is always larger than the outer one by a factor of around 2, except at large $N$, when relatively few points are used for the fit and the ratio exceeds 3. This could indicate that the inner and the outer sides correspond to different ages, which would suggest that the Tirela/Klumpkea family is the result of more than one collisional event. 
However, this hypothesis can hardly be confirmed in the future, as the inner side has probably been substantially eroded by the 11:5 and 13:6 mean motion resonances with Jupiter (centered at 3.075 and 3.105 AU, respectively). 

To convert the slopes into ages, it is necessary to determine a Yarkovsky calibration, represented by the value of the drift $da/dt$ for a family member with a unit diameter and spin axis obliquity equal to 0° for the inner side and to 180° for the outer side \citep{spoto-2015}. \cite{milani-2014} and \cite{spoto-2015} constrain the Yarkovsky calibration using (101955) Bennu as a benchmark asteroid. However, (101955) Bennu is a B- type whose spectral properties are very different from the two families considered in this work. We therefore selected as reference asteroid (99942) Apophis, which instead is an S- type and thus closer to Tirela/Klumpkea and Watsonia. \cite{fenucci-2024} report a Yarkovsky drift of $-13.26$ AU/Myr for (99942) Apophis measured with $S/N \approx 123$. The scaling formula thus becomes:

\begin{equation}
\frac{da}{dt} = \left( \frac{da}{dt} \right)_A \ \frac{\sqrt{a}_A \ (1-e_A^2)}{\sqrt{a} \ (1-e^2)} \ \frac{D_A}{D} \ \frac{\widetilde{\rho_A}}{\widetilde{\rho}} \ \frac{\cos(\phi)}{\cos(\phi_A)} \ \frac{1-\rho_V}{1-\rho_{V, A}}
\label{EqYarkCalibration}
,\end{equation}

where $a$, $e$, $\widetilde{\rho}$ and $\rho_V$ are respectively the semi-major axis, the eccentricity, the density and the albedo representative for the family. Then, $\cos(\phi)$ is the spin axis obliquity and it is equal to $\pm 1$ depending if the inner or the outer side is considered. $D = 1$ km is not the diameter of a real asteroid, but the reference value corresponding to the inverse slope. Also, $a_A$ and $e_A$ are the semi-major axis and the eccentricity of (99942) Apophis \citep{moskovitz-2022}, $D_A$ is its diameter \citep{brozovic-2018}, $\rho_{V, A}$, and $\cos(\phi_A)$ are its albedo and its spin axis obliquity \citep{berthier-2023}.\\
The densities $\widetilde{\rho}$ and $\widetilde{\rho_A}$ are both unknown. For (99942) Apophis, we assume the reference density for S- types reported in \cite{carry-2012}, $2.70 \pm 0.69$ g/cm$^3$. For the family, we could have assumed the reference density for L- types reported in the same work. However, this value is quite inaccurate due to the limited number of large L- types with precise measurements of mass and diameter. In addition, \cite{sunshine-2008} and \cite{mahlke-2023} suggested that Barbarians are linked to CO and CV meteorites, which are, in turn, related to C- types. We thus assumed for the family the reference density for the C- class reported in \cite{carry-2012}, $1.41 \pm 0.69$ g/cm$^3$. As in \cite{spoto-2015}, we assume the Yarkovsky calibration to be affected by a relative uncertainty of 0.3. We thus obtain a  result of $da/dt = (5.86 \ \pm \ 1.76) \ 10^{-4}$ AU/Myr for the family.
\newline
\indent \newtext{Figure \ref{AgesTirela} reports the ages for the inner side ($200 \ \pm  97$ Myr) and the outer side ($467 \ \pm  184$ Myr) compared to the values reported in Mi16. The age found by our work for the outer side is larger than the one for the inner side, but they are compatible within the error bars. The difference is probably due to a poor accuracy of the fit in the inner side, due to the limited extension of this region. In Fig. \ref{AgesTirela}, we only give the case for $N=13$ (see Fig. \ref{FitVShapeTirela}), which is the one affected by the smallest uncertainties. For every other value of $N,$ the ages for the two sides are always compatible among them, except for $N=14$, which is very likely the result of an inaccurate fit of the inner side.}\\
The ages for the outer side are all compatible with Mi16, while the ages for the inner side are systematically below it. This may suggest that the Tirela/Klumpkea family is younger than previously thought. However, the ages are affected by the choice of the reference asteroid and the values of the densities used in Eq. (\ref{EqYarkCalibration}), therefore this result cannot be considered as definitive. Finally, our age estimation is compatible with \citet{broz-2013}, who exploited N-body numerical simulations and ultimately found the Tirela/Klumpkea family to be younger than 1 Gyr.

\section{The difficult case of Watsonia} \label{SectionWatsonia}

Watsonia is an asteroid family first identified by \newtext{\cite{novakovic-2011}}, who applied the classical HCM to high-inclination main belt asteroids. Watsonia is located in the middle of the main belt, with 2.70 AU $< a <$ 2.83 AU, and it is characterized by low to moderate orbital eccentricities, $0.10 < e < 0.15$, and high inclinations, $16.5^{\circ} < i < 18.0^{\circ} $. Watsonia lies in a region of the proper elements with a low density of asteroids, where the difference in spectral type between the background and the family members is less marked than in the Tirela/Klumpkea case, thus reducing the efficacy of separating the family members from the background using \gaia data alone. Watsonia is a very small family, with just a few tens of members reported in the literature (e.g., the HCM identified only 80 members). \cite{tsirvoulis-2014} tried to constrain the age of the family using the V-shape method, but because of the low number of family members, these authors could not achieve a precise estimation and they report an approximate age of $\sim$1~Gyr.

\begin{figure}
   \centering
   \includegraphics[width=\hsize]{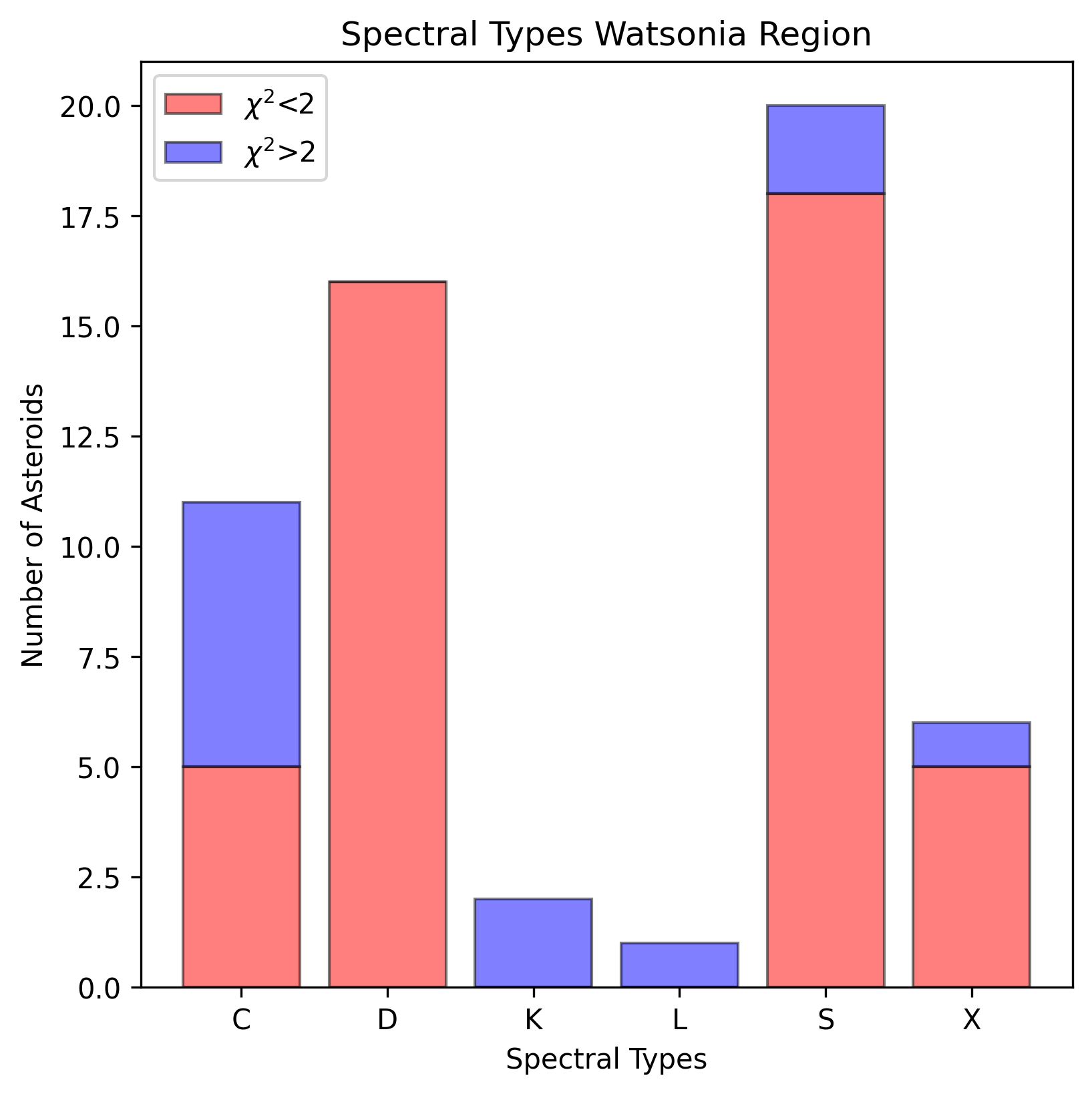}
      \caption{\small Spectral types of the objects observed by \gaia in the Watsonia region. Blue are objects with $\chi^2 > 2$, red are objects with $\chi^2 < 2$.
              }
         \label{TaxonomyWatsonia}
   \end{figure}

\begin{figure}
   \centering
   \includegraphics[width=\hsize]{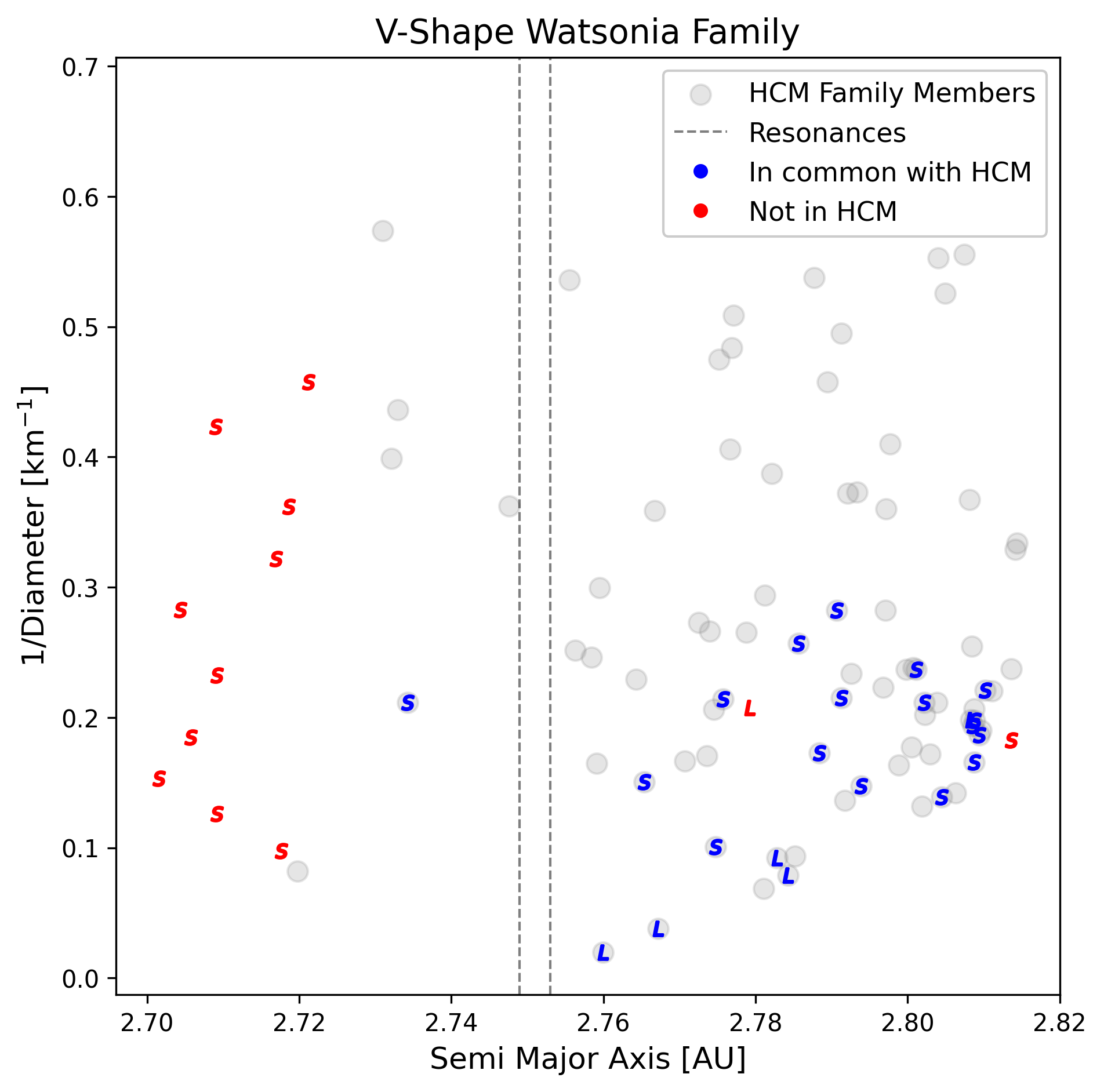}
      \caption{\small Distribution in the $(a, 1/D)$ plane of the Watsonia family members, marked by the corresponding taxonomy class letters. In blue, we show objects in common between our work and the HCM. In red, we show objects that are not identified by the HCM. The grey circles indicate all the family members identified by the HCM. 
              }
         \label{VShapeWatsonia}
   \end{figure}

\begin{figure*}
   \centering
   \includegraphics[width=\textwidth]{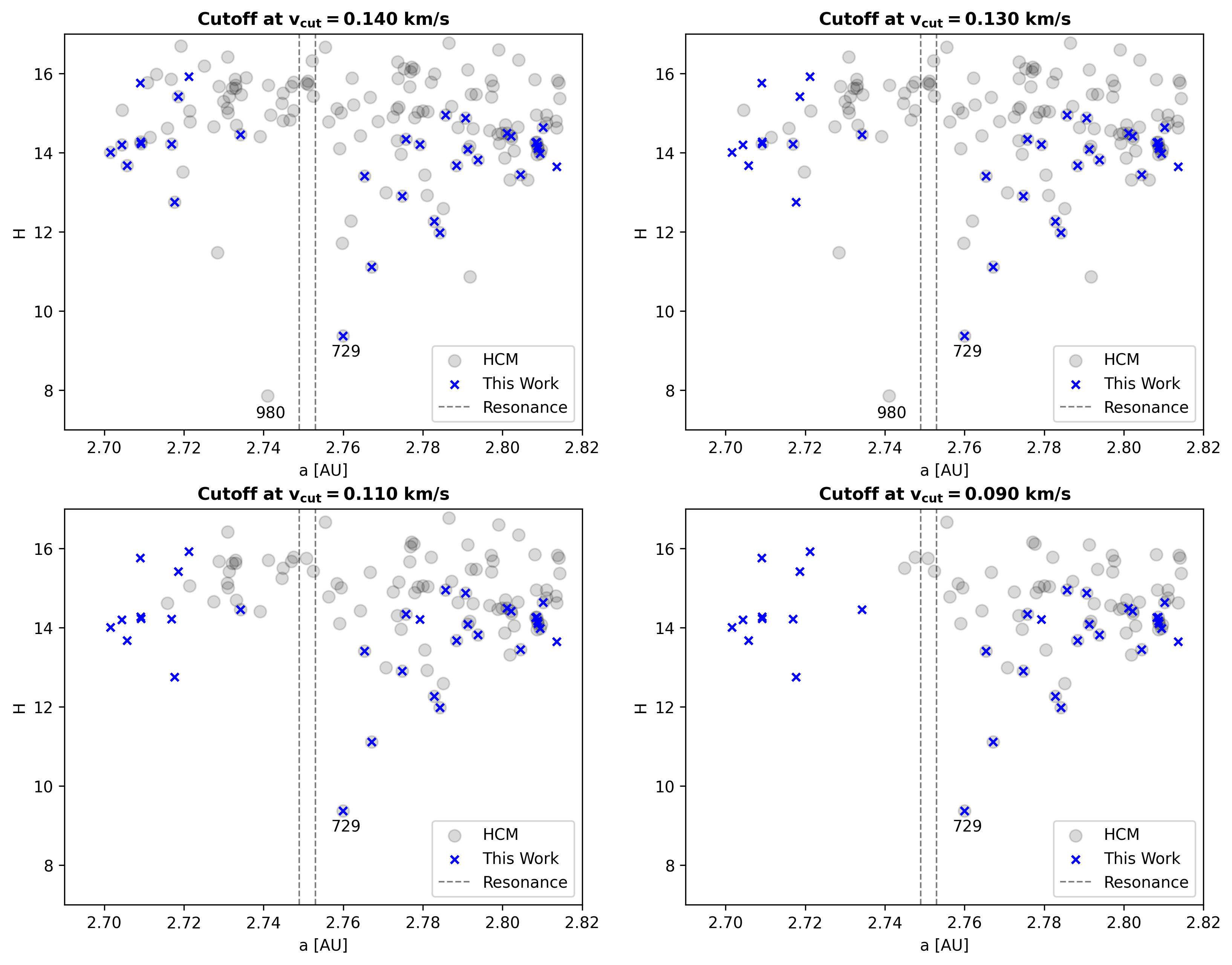}
      \caption{\small Comparison between the HCM membership (grey circles) and our work (blue crosses) at different cutoff velocities in the ($a$, $H$) plane. The numbers mark the positions of (729) Watsonia and (980) Anacostia.
              }
         \label{WatsoniaComparisonNewHCM}
   \end{figure*}
   
To identify the Watsonia family, the same procedure implemented for Tirela/Klumpkea was followed. First, all asteroids contained inside a region centered on (729)~Watsonia with 2.70 AU $< a <$ 2.83 AU, $ 0.105 < e < 0.150,$ and $0.29 < \sin(i) < 0.31$ are considered. Again, the proper elements and the absolute magnitudes are retrieved from the Asteroid Family Portal, and the albedos from the NEOWISE survey. Only 227 asteroids are found within this region, 56 of which present a spectrum in \gaia DR3, while only 6 of them had a spectrum previously published from other sources. Among these, the three brightest family members found by the HCM are selected to create a family template spectrum (Fig.~\ref{TemplateWatsonia}). 

The same $\chi^2$ technique (Eq.~\ref{EqChiSquare}) is then applied to quantify the differences between the family template and the objects observed by \gaia. The distributions in the $(\chi^2, \rho_V)$ and $(a, 1/H)$ planes for the objects within the Watsonia region are shown in Fig.~\ref{ChiSquareAlbedoWatsonia}. The points are divided into four subgroups following the approach used for Tirela/Klumpkea. The small number of objects however makes the result less sharply defined.

The spectral types resulting from the color taxonomy are reported in Fig.~\ref{TaxonomyWatsonia}.  
Despite the small statistics, there is a quite good agreement between spectral types and $\chi^2$, with background objects with $\chi^2 > 2$ classified into the C- class, while almost all D- and S- types are associated with $\chi^2 < 2$.\\
Quite surprisingly only one object is classified as L- type, with even (729)~Watsonia itself included in the X- class. The only L- type, which in addition has $\chi^2 > 2$, is (980)~Anacostia, the lowest point around 2.74 AU in the right panel of Fig. \ref{ChiSquareAlbedoWatsonia}, whose puzzling proximity to the family has already been discussed in the literature. \cite{cellino-2014} in particular mention that (980) Anacostia and (729) Watsonia share similar semi-major axes and inclinations and they present similarities in their physical properties, such as being both Barbarians (\citealt{gilhutton-2008}, \citealt{gilhutton-2014}). They speculated that the two objects might have a similar origin from an ancient disruption of a large Barbarian parent body. This is also suggested by the presence of other large Barbarians close to (729) Watsonia and (980) Anacostia in the phase space of proper elements, such as (387) Acquitania and (599) Luisa. The smallest fragments created in this event would have been removed by the Yarkovsky drift, while the largest fragments had time to slowly evolve in the proper elements, assuming their actual positions. In this picture, Watsonia would be a second-generation family formed from the disruption of one of the first-generation large fragments. \\
Our V-plot shows that (980)~Anacostia seems to lie on the alignment of a poorly defined right edge of the family. However, we find that in terms of the reflectance spectra obtained by \gaia, (980) Anacostia slightly differs from the Watsonia family, as reported in Fig. \ref{TemplateWatsonia}. As we will see further on, the analysis of the HCM will not be conclusive for (980)~Anacostia, although it provides other interesting clues.

Analyzing the objects with $\chi^2 < 2$ and initially classified as C-, D- and X- types, we find that about half of the C- and X- types and two-thirds of the D- types receive S- or L- as second classification. For all of them, the probabilities of the first and second classification are very similar and the spectra closely resemble the family template spectrum. We thus decided to consider as family members all S- types with $\chi^2 < 2$ and the C-, D- and X- types with $\chi^2 < 2$, having S or L as the second most probable classification. A total of 34 family members are retrieved from the \gaia spectra and the color taxonomy. The complete family membership for the Watsonia family is reported in the Zenodo repository \footnote{\url{https://doi.org/10.5281/zenodo.12704069}}.

The absolute magnitudes, $H,$ of the family members are converted into diameters, $D,$ from the albedo, either extracted from NEOWISE or (when unavailable) adopting the average albedo of the family, $\rho_v=0.16 \pm 0.06$. 
The distribution of family members in the $(a, 1/D)$ plane is reported in Fig.~\ref{VShapeWatsonia}, where every object is marked by the corresponding taxonomy class letter. For the objects that did not receive S as the most probable class, but that were still included in the family, their second most likely classification is reported instead. The vertical dashed lines mark the position of a resonance between 2.749 AU and 2.753 AU, within which \gaia did not detect any object. The resonance lies very close to the position of (729)~Watsonia at 2.76 AU. \\
S- types are distributed all over the area, while the few L- types are mainly found at $D > 10$ km. The three asteroids used to create the template family spectrum are indeed the three largest L- types.\\
Despite a limited statistic, the V-shape is recognizable especially in the outer side. In addition, almost all objects at the right of the resonance are included in the family by the HCM. This is not a coincidence, since most family members identified by the HCM lie in this region, as shown by the grey circles in Fig. \ref{VShapeWatsonia}.
The most striking feature in Fig.~\ref{VShapeWatsonia} is the appearance in the family of a set of objects at the left of the resonance (marked in red), that are not identified by the HCM. In percentage, our analysis identifies more objects in the inner side than the HCM, according to which the family seems to be one-sided. This however could be the result of the limit cutoff velocity chosen by AstDyS to define the family. 

Figure \ref{WatsoniaComparisonNewHCM} reports the comparison between the family members identified in this work and the membership obtained by our HCM at different cutoff velocities $v_{cut}$ in the ($a$, $H$) plane. At the smallest cutoff velocity, $v_{cut} = 0.090$ km/s, the family membership retrieved by our HCM is very similar to that of AstDyS, with most family members lying at the right of the resonance. Increasing $v_{cut}$ the overall shape of the outer side remains unchanged, while the HCM links more and more objects to the inner side at the left of the resonance. At $v_{cut} = 0.140$ km/s almost all family members retrieved by our work are linked to the family by the HCM, while the slopes of the inner and outer sides seem to be qualitatively compatible. Interestingly, for this same limit, (980) Anacostia is also included in the family by the HCM.

\begin{figure}
   \centering
   \includegraphics[width=\hsize]{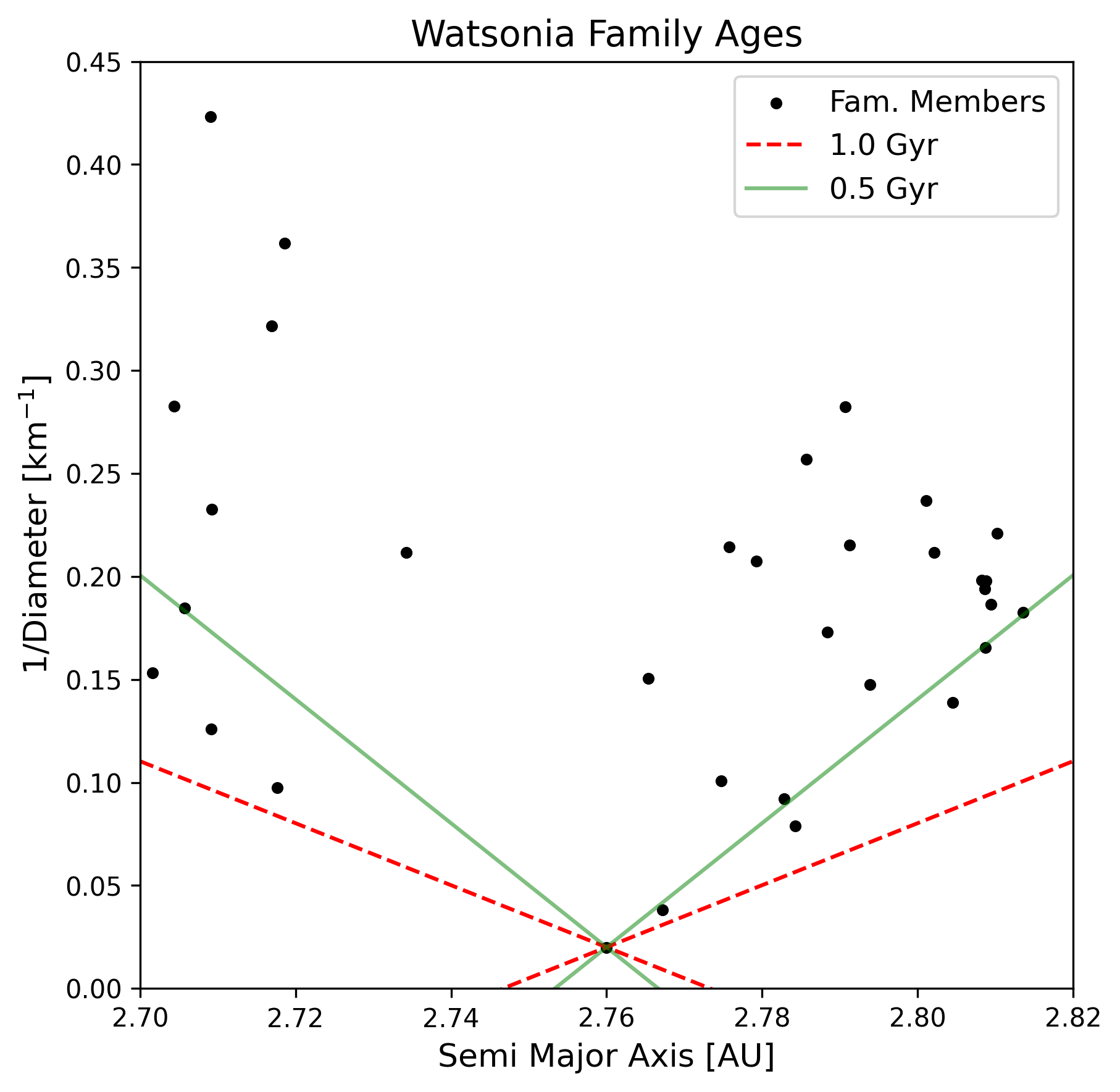}
      \caption{\small Family members in the $(a, 1/D)$ plane fitted by V-shapes corresponding to different ages. The green continuous line corresponds to 0.5 Gyr and the red dashed line to 1.0 Gyr.}
         \label{AgesWatsonia}
   \end{figure}

The scarcity of members in our limited data set prevents to have solid statistics for the application of the full V-shape fitting method. To have an approximate estimation of the age of Watsonia, we overplot the tentative V-shape boundaries corresponding to specific ages (Fig.~\ref{AgesWatsonia}), calibrated from Eq. (\ref{EqYarkCalibration}) by considering the semi-major axis, eccentricity, and albedo of (729) Watsonia and assuming the reference C- type density reported in \cite{carry-2012}, following the link to the CV and CO meteorites suggested by \cite{mahlke-2023} and \cite{sunshine-2008}. 
The age of the Watsonia family seems to lie somewhere in between the V-shapes corresponding to 0.5 Gyr and 1.0 Gyr, a value in agreement with \cite{tsirvoulis-2014}. However, due to the limited number of family members and the large uncertainty on the Yarkovsky calibration, this result is affected by large errors and thus \newtext{can only be considered as a rough estimate.}

\section{Conclusions} \label{Conclusion}

Our classification of \gaia spectra, not using external etalon templates, is successful in separating the main spectral types. Our analysis of specific families shows that some expected ambiguities remain, especially when trying to push the analysis toward family members of smaller sizes (i.e., with fainter magnitudes and noisier spectra). \\
The two Barbarian families that we analyze here have slightly different reflectance spectra, as shown by the distribution of taxonomy. This is consistent with the differences already found in the past in polarimetric parameters and when matching their members to meteorite spectra. \citet{devogele-2018} found that Watsonia, despite having anomalous polarimetry common to other Barbarians, requires less extreme enrichment in CAIs to be modeled. These two families also differ in their position in the belt and in the number of members bright enough to be accessible to \gaia (Tirela/Klumpkea being the richest). 

The case of Tirela/Klumpkea is a significant benchmark for spectral classification and family membership, in fact its members exhibit an albedo markedly higher than the average background dominating the outer belt, where it is situated. In this case, the comparison of albedo with \gaia spectra and HCM membership allows us to identify interlopers among some large family members. \\
Smaller members are also recovered by spectral similarity. Although in general many of them do not have albedo measurements available, they qualitatively scatter on the osculating elements plane on regions compatible with the evolved family halo. 
A simulation of the dynamical evolution of the family, outside the scope of this article, would be useful to further consolidate this result. \\
Interestingly, the new members seem to better define the inner slope of the V-shape of the family, resulting in a younger age (200 Myr) with respect to \citet{milani-2016}. The outer edge also appears younger but not by a significant margin. Our determination of family membership eliminates the large interlopers included by the HCM that polluted the V-shape and resulted in a very peculiar size distribution.

The Watsonia family, as expected, is a more difficult case than Tirela/Klumpkea, but it still represents an interesting application of our approach. The limited number of objects around the family and the less marked difference between the spectral types of the family members and the background make the identification by \gaia data alone more difficult. However, \gaia observed 56 objects inside the Watsonia region, thus representing a huge improvement compared to the literature, where only six spectroscopic observations for these objects have been reported .\\
Despite the fact that the V-shape boundaries remain poorly determined, we are capable of identifying several new candidate family members that had not been included in previous HCM-based classifications, drafting the structure of the inner part of the V-shape. Interestingly, our new run of the HCM shows that at the same cutoff velocity required to include the inner members, (980)~Anacostia appears in the membership. This evidence strengthens the physical and dynamical links among different large Barbarians in the area, which possibly stands as a sparse signature of an ancient breakup from which the Watsonia parent body was originated, previously suggested by \citet{cellino-2014} and others. A similar consideration could be applied to the large asteroids that we find to be compatible to Tirela/Klumpkea in terms of spectra, but that lie totally outside the core of the V-shape (left panel of Fig.~\ref{VShapeTirela}). Their presence and proximity may also indicate more ancient break-ups of large L-type objects. 

The reconstruction of a ''genealogy'' leading to the currently observed families would be a major milestone in understanding the origin and nature of L-type families and Barbarians. We have shown how asteroid reflectance spectra by \gaia can be used to work in this direction. Future releases from this same mission will offer a larger sample of objects to study, with improved data quality. Its exploitation, jointly with data coming from other surveys, can be the key to future advances in our understanding of the origin and history of the asteroid population.

\begin{acknowledgements}
This work presents results from the European Space Agency (ESA) space mission \gaia. \gaia data are being processed by the \gaia Data Processing and Analysis Consortium (DPAC). Funding for the DPAC is provided by national institutions, in particular, the institutions participating in the \gaia Multilateral Agreement (MLA). The \gaia mission website is https://www.cosmos.esa.int/\gaia. The \gaia archive website is https://archives.esac.esa.int/Gaia.

RB Doctoral contract is funded by Universit\`e de la C\^ote d'Azur.

This project was financed in part by the French Programme National de Planetologie, and by the BQR program of Observatoire de la C\^ote d'Azur. 

We made use of the software products: SsODNet VO service of IMCCE, Observatoire de Paris, and the associated {\tt rocks} library\footnote{\url{https://github.com/maxmahlke/rocks}} \citep{berthier2022ssodnet}; Astropy, a community-developed core Python package for Astronomy \citep{0astropy2013, 1astropy2018, 2astropy2022}; Matplotlib \citep{matplotlib_Hunter:2007}.
\end{acknowledgements}

%
%
\bibliographystyle{aa}
\bibliography{references}

\end{document}